\documentclass[pra,letterpaper,aps,10pt,superscriptaddress,twocolumn,floatfix,showpacs]{revtex4-1}
\usepackage{graphicx}
\usepackage{amsmath}
\usepackage{amsfonts}
\usepackage{amssymb}
\usepackage{epsfig}
\usepackage{epstopdf}
\DeclareGraphicsExtensions{.pdf,.eps,.png,.jpg,.mps}
\usepackage[pdftex]{color}
\usepackage{amsmath,graphicx,amssymb,braket,xcolor,subfigure,upgreek}
\usepackage[colorlinks, linkcolor=blue, citecolor=blue, urlcolor=blue, breaklinks=true]{hyperref}
\usepackage{microtype}
\usepackage{bbm}
\usepackage{xcolor}
\usepackage{braket}
\usepackage{amsmath,amssymb,amsfonts,latexsym,color,dcolumn,bm,amsbsy}
\usepackage[english]{babel}
\usepackage{graphicx}
\usepackage[utf8]{inputenc}
\usepackage{textcomp}
\usepackage{braket}
\usepackage{bbm}
\usepackage{float}
\usepackage{mathtools}
\usepackage{soul}

\makeatletter
\newsavebox\myboxA
\newsavebox\myboxB
\newlength\mylenA

\newcommand*\xoverline[2][0.75]{%
	\sbox{\myboxA}{$\m@th#2$}%
	\setbox\myboxB\null% Phantom box
	\ht\myboxB=\ht\myboxA%
	\dp\myboxB=\dp\myboxA%
	\wd\myboxB=#1\wd\myboxA% Scale phantom
	\sbox\myboxB{$\m@th\overline{\copy\myboxB}$}%  Overlined phantom
	\setlength\mylenA{\the\wd\myboxA}%   calc width diff
	\addtolength\mylenA{-\the\wd\myboxB}%
	\ifdim\wd\myboxB<\wd\myboxA%
	\rlap{\hskip 0.5\mylenA\usebox\myboxB}{\usebox\myboxA}%
	\else
	\hskip -0.5\mylenA\rlap{\usebox\myboxA}{\hskip 0.5\mylenA\usebox\myboxB}%
	\fi}
\makeatother
\definecolor{MDBlue}{rgb}{0.1, 0.1, 0.44}

\begin{document}
	
	\title{Nonlinear opto-vibronics in molecular systems}
\author{Q. Zhang}
\affiliation{Max Planck Institute for the Science of Light,
	D-91058 Erlangen, Germany}
\author{M. Asjad}
\affiliation{Department of Mathematics, Khalifa University, Abu Dhabi 127788, United Arab Emirates}
\author{M. Reitz}
\affiliation{Department of Chemistry and Biochemistry, University of California San Diego, La Jolla, California 92093, USA}
\author{C. Sommer}
\affiliation{Alpine Quantum Technologies GmbH, 6020 Innsbruck, Austria}
\author{B. Gurlek}
\affiliation{Max Planck Institute for the Structure and Dynamics of Matter and Center
for Free-Electron Laser Science, 22761 Hamburg,
Germany}
\author{C. Genes}
\affiliation{Max Planck Institute for the Science of Light,
	D-91058 Erlangen, Germany}
\affiliation{Department of Physics, Friedrich-Alexander-Universit\"{a}t Erlangen-N{\"u}rnberg (FAU),
	D-91058 Erlangen, Germany}
\date{\today}

\begin{abstract}
	We analytically tackle opto-vibronic interactions in molecular systems driven by either classical or quantum light fields. In particular, we examine a simple model of molecules with two relevant electronic levels, characterized by potential landscapes with different positions of minima along the internuclear coordinate and of varying curvatures. Such systems exhibit an electron-vibron interaction, which can be comprised of linear and quadratic terms in the vibrational displacement. By employing a combination of conditional displacement and squeezing operators, we present analytical expressions based on a quantum Langevin equations approach, to describe the emission and absorption spectra of such nonlinear molecular systems. Furthermore, we examine the imprint of the quadratic interactions onto the transmission properties of a cavity-molecule system within the collective strong coupling regime of cavity quantum electrodynamics.
\end{abstract}

\maketitle

\section{Introduction}

Opto-vibrational interactions in molecular systems occur in an indirect fashion as light couples to electronic transitions, which are in turn coupled to the vibrations of nuclei~\cite{neuman2018origin,clark1986resonance,smith2005chapter4,riede2018organic}. A standard description of electron-vibron interactions, under the Born-Oppenheimer approximation, is given by the Holstein Hamiltonian \cite{holstein1959study,Spano2016excitons} which is a spin-boson model linear in the vibrational displacement.
%%%%%%%%%%%%%%%%%%%%%%%%%
%%Figure 1
%%%%%%%%%%%%%%%%%%%%%%%%%
\begin{figure}[b]
	\centering
	\includegraphics[width=0.99\columnwidth]{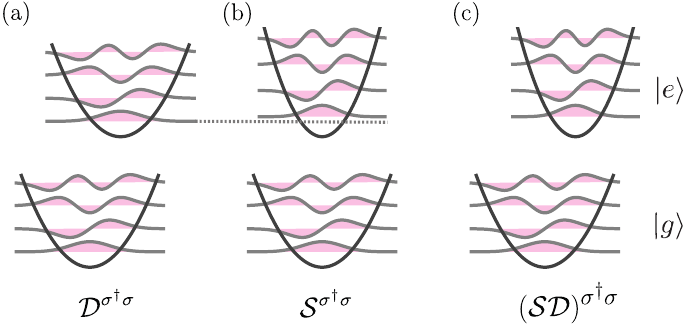}
	\caption{ (a) Standard scenario where the excited state potential landscape is a copy of the ground state landscape slightly shifted. Electronic excitation is accompanied by the action of a conditional displacement operator $\mathcal{D}^{\sigma^\dagger \sigma}$, where $\sigma$ is the ladder operator from the excited to the ground electronic state. (b) Scenario with unshifted potentials with different curvatures. Electronic excitation is accompanied by a conditional squeezing operation $\mathcal{S}^{\sigma^\dagger \sigma}$. (c) Combined model where electronic excitation leads to a displacing and squeezing operation.}
	\label{fig1}
\end{figure}
%%%%%%%%%%%%%%%%%%%%%%%%%
%%%%%%%%%%%%%%%%%%%%%%%%%
Some analytical treatments based on quantum Langevin equations (QLEs)~\cite{reitz2019langevin,reitz2020molecule,zhang2023multidimensional,kansanen2021polariton,kansanen2019theory} have been shown to provide approximate analytical results for this model for a large number of vibrational modes and in the presence of fast vibrational relaxation typically occurring in both bulk~\cite{reitz2020molecule,keeling2020bose} and solvent environments~\cite{gilmore2005spin}. Similar methods have been used in cavity optomechanics~\cite{aspelmeyer2014cavity,aspelmeyer2014cavity_RMP}, where cavity-confined quantum light modes are coupled to macroscopic oscillators via the radiation pressure Hamiltonian, to study the strong photon-phonon coupling regime~\cite{rabl2011photon,nunnenkamp2011single}.

Such theoretical treatments are based on a polaron transformation which allows for the diagonalization of the bare Holstein Hamiltonian~\cite{jang2022partially}. This can be understood as a conditional displacement operation, where the electronic state dictates whether or not a displacement in the vibrational subspace should be performed. In consequence, when a photon excites an electronic transition between two copies of the same harmonic potential landscape slightly shifted (see Fig.~\ref{fig1}(a), the vibrational state is excited to a coherent state. The underlying assumption here is however that the potential landscapes are identical. In reality it can happen that the curvatures of the two potential energy surface are different, as illustrated in Fig.~\ref{fig1}(b): an electronic transition will then be accompanied by a squeezing of the vibrational wave-packet. In such a case the polaron transformation is modified by an operation involving a conditional squeezing operator. Most generally, one can imagine the situation depicted in Fig.~\ref{fig1}(c) where the proper diagonalizing transformation involves a conditional displacement followed by squeezing. In optomechanics, this corresponds to a quadratic photon-phonon interaction~\cite{liao2013photon}.

We provide here an analytical treatment based on a set of QLEs for effective spin operators dressed by vibrations, which can be solved under some approximations to provide information about emission and absorption spectra. Additionally, we investigate the transmission properties of an optical cavity within the strong coupling regimes of cavity quantum electrodynamics. By studying the interaction between the molecular systems and the cavity, we gain insight into the nature of light-matter interactions in these complex environments.

The paper is organized as follows: in Sec.~\ref{Model} we introduce the modified Holstein model obtained from first principle derivations of the electron-vibration coupling for a scenario depicted in Fig.~\ref{fig1}(c). Our analytical treatment is based on a set of simplified QLEs for vibrations and electronic degrees of freedom as derived in Sec.~\ref{QLE}. We proceed with solving the QLEs under the approximation of weak excitation of the upper electronic state to obtain absorption and emission spectra under illumination with classical light. Finally, in Sec.~\ref{sec:Molecular-Cavity} we add a quantum confined light field coupled to the electronic transition via the Tavis-Cummings Hamiltonian and derive the transmission profile of the cavity in the weak and strong coupling regimes of light-matter interactions.
	
	\section{The modified Holstein model}
	\label{Model}
	
	We consider a molecule with two relevant electronic states denoted by $\ket{g}$ and $\ket{e}$ for ground and excited, respectively. Transitions between these two states are characterized by Pauli lowering operators $\sigma=\ket{g}\bra{e}$  and its corresponding Hermitian conjugate. As illustrated in Fig.~\ref{fig1}(c), the ground/excited potential landscapes are assumed to have a parabolic shape, with the minima of these two potential landscapes separated by $R_{eg}$ and with different curvatures, thus having different vibrational frequencies: $\nu_g$ for the electronic ground state and $\nu_e$ for the electronic excited state. The Hamiltonian describing the molecular system can be expressed as ($\hbar=1$) 	
	\begin{equation}
		\begin{aligned}
			\mathcal{H}=\mathcal{V}_e(\hat R,\hat P)\sigma^\dagger\sigma+\mathcal{V}_g(\hat R,\hat P)\sigma\sigma^\dagger,
		\end{aligned}\label{eq:Ham_Orig}
	\end{equation}
where $\mathcal{V}_e$ and $\mathcal{V}_g$ denote the potential landscapes in the electronic excited and ground state, respectively, defined onto the direction of the nuclear coordinate as
\begin{subequations}
	\begin{align}
		\mathcal{V}_e(\hat R,\hat P) & =\omega_e+\frac{\hat P^2}{2\mu}+\frac{1}{2}\mu\nu_{e}^2(\hat R-R_\mathrm{eg})^2,\\
		\mathcal{V}_g(\hat R,\hat P) &= \omega_g+\frac{\hat P^2}{2\mu}+\frac{1}{2}\mu\nu_{g}^2\hat R^2,
	\end{align}
\end{subequations}
with the reduced mass $\mu$, the momentum operator and position operators $\hat P$ and $\hat R$ satisfying the commutation relation $[\hat R,\hat P]=i$. Notice that the matrix elements of the Hamiltonian $\mathcal{H}$ can be written in a basis formed by $\{\ket{g;m_g}=\ket{g}\otimes\ket{m_g},\ket{e;m_e}=\ket{e}\otimes\ket{m_e}\}$, where the Fock states $\ket{m_e}$ and $\ket{m_g}$, respectively, refer to the eigenstates of the vibrational Hamiltonian part contained in $\mathcal{V}_g(\hat R,\hat P)$ and $\mathcal{V}_e(\hat R,\hat P)$, respectively.
	
However, one can express the quadratures in terms of creation $b^\dagger=(\hat R/{R_{\mathrm{zpm}}}-iR_{\mathrm{zpm}}\hat P)/\sqrt{2}$ and annihilation $b=(\hat R/{R_{\mathrm{zpm}}}+iR_{\mathrm{zpm}}\hat P)/\sqrt{2}$ operators. The operators fulfill the following commutation $\left[b, b^\dagger\right]=1$ and the zero-point motion is defined as $R_\mathrm{zpm}=1/\sqrt{2\mu\nu_{g}}$. Notice that the definition of this bosonic operator is performed with respect to the ground state such that it diagonalizes the ground state vibrational problem. The Hamiltonian in Eq.~\eqref{eq:Ham_Orig} can now be written as
	\begin{equation}
		\begin{aligned}
				\mathcal{H} = &\nu_g b^\dagger b +\omega_{0}\sigma^\dagger  \sigma+\lambda_1 \nu_g  (b + b^\dagger )\sigma^\dagger  \sigma\\
					&+\lambda_2 \nu_g (b + b^\dagger )^2\sigma^\dagger  \sigma.
		\end{aligned}
	\label{eq:Ham}
	\end{equation}
The linear coupling parameter results from the mismatch in the positions of the minima $\lambda_1=  -\mu\nu_{e}^{2}R_\mathrm{eg}R_{\mathrm{zpm}}/\nu_{g}$ while the quadratic coupling parameter is proportional to the relative change in vibrational frequencies $\lambda_2=\left(\nu_{e}^{2}-\nu_{g}^{2}\right)/\left(4\nu_{g}^{2}\right)$. The bare electronic frequency splitting is modified by the vibronic coupling $\omega_{0}=\omega_e-\omega_g+\lambda_1^2\nu_g^3/\nu_e^2$.

\begin{figure}[t]
	\centering
	\includegraphics[width=0.90\columnwidth]{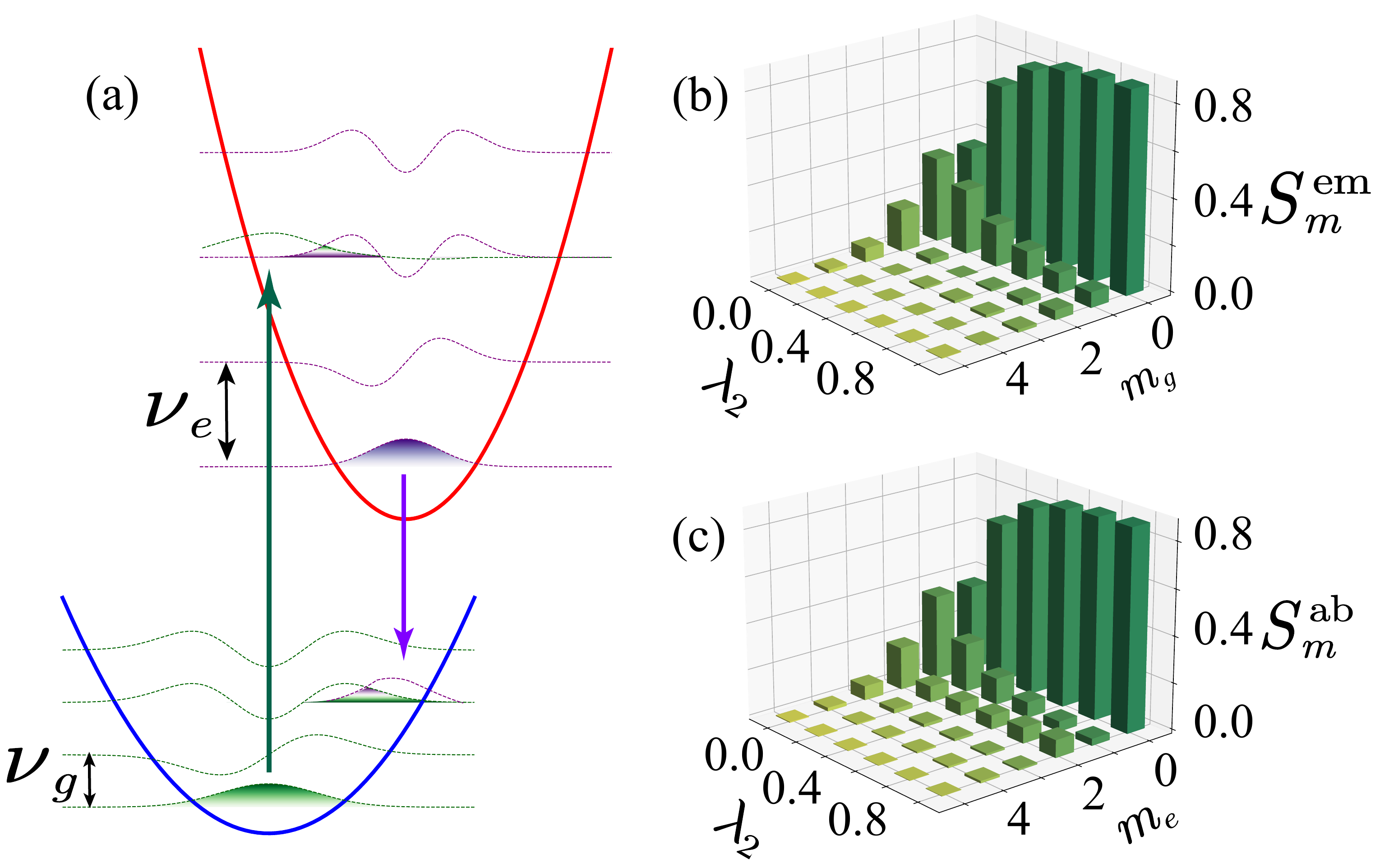}
	\caption{(a) Schematic diagram of a molecular system exhibiting two parabolic electronic potential surfaces, slightly shifted and with different curvatures quantified by the vibrational frequencies $\nu_g$ and $\nu_e$.  Histogram of the vibrational state occupancy in the electronic ground state upon emission from $\ket{e,0_e}$ in (b) and in the electronic excited state upon external drive from the $\ket{g;0_g}$ state in (c) for various values of $\lambda_2$ at fixed $\lambda_{1}=1$.}
	\label{fig2}
\end{figure}
%%%%%%%%%%%%%%%%%%%%%%%%%
%%%%%%%%%%%%%%%%%%%%%%%%%
However, it is more convenient to use a single basis formulation where only the eigenstates of the harmonic oscillator in the ground state are considered, i.e., the eigenstates of $\nu_g b^\dagger b$ denoted by $\{\ket{m_g}\}$. To this end, one can take the level-dependent unitary transformation $\tilde{\mathcal{H}}=\mathcal{U}^\dagger \mathcal{H} \mathcal{U}$ with
	\begin{equation} \mathcal{U}=\mathcal{D}(r_d)^{\sigma^\dagger\sigma}\mathcal{S}(r_s)^{\sigma^\dagger\sigma}=\sigma\sigma^\dagger+\mathcal{D}(r_d)\mathcal{S}(r_s)\sigma^\dagger\sigma.
	\end{equation}
The definitions of the displacement and squeezing operators are the standard ones employed in quantum optics
\begin{equation}
	\begin{aligned}
		\mathcal{D}(r_d)=e^{r_d(b^\dagger-b)}\quad\mathrm{and}\quad \mathcal{S}(r_s)= e^{\frac{1}{2}r_s\left(b^2-b^{\dagger2}\right)}
	\end{aligned}
\end{equation}
which employ the following displacement $r_d$ and squeezing $r_s$ parameters defined as
\begin{equation}
	r_d=-\lambda_1\frac{\nu_g^2}{\nu_e^2} \quad \mathrm{and} \quad r_s=\frac{1}{2}\left(\ln\nu_{e}-\ln\nu_g\right).
\end{equation}
Finally, the Hamiltonian is expressed in diagonal form
\begin{equation}
	\tilde{\mathcal{H}}=\nu_gb^\dagger b\sigma\sigma^\dagger+\left(\nu_eb^\dagger b+\omega_{0 0}\right) \sigma^\dagger\sigma,
\end{equation}
where  the effective frequency $\omega_{00}=\omega_e-\omega_g+(\nu_{e}-\nu_{g})/2$ relates to the zero-phonon line.

This is nothing more than a generalized polaron transformation where the electronic coherence operator $\sigma$ is \textit{dressed} by the vibrational modes as $\sigma  \mathcal{D} (r_d) \mathcal{S} (r_s)$ via both a displacement and a squeezing operation. This offers a recipe to obtain the intensity of vibronic transitions in the emission and absorption processes. Assuming the molecule initially in the excited state with zero vibrations $\ket{e;0_e}$, the  probability of ending up in the state $\ket{g;m_g}$ is governed by the overlap between the two vibrational wave functions [see Fig.~\ref{fig2}(a)] as
%can be computed as matrix elements $\bra{n}\mathcal{S} (r_s) \mathcal{D} (r_d) \ket{0}$. Thus the probability of ending up in vibrational level $n$ upon excitation can be expressed as
\begin{equation}
	\begin{aligned}
		S_m^{\mathrm{em}}=&\left|\braket{m_g\vert 0_e}\right|^2=\left|\bra{m_g}\mathcal{S} (r_s) \mathcal{D} (r_d) \ket{0_g}\right|^2\\
		=&\frac{e^{-r_d^2\alpha}}{\cosh(r_s)}\left[H_{m}\left(\frac{\alpha r_d}{2\sqrt{\beta}}\right)\right]^2\frac{\beta^{m}}{m!},
	\end{aligned}
\end{equation}
where $H_m(x)$ are Hermite polynomials, $\alpha=\tanh r_s+1
$, and $\beta=\left(\tanh r_s\right)/2$.  Similarly, we can find the absorption  probability amplitude for the absorption transition $\ket{g;0_g}\rightarrow\ket{e;m_e}$ via the Hermitian adjoint operator $\sigma^\dagger\mathcal{S}^\dagger(r_s)\mathcal{D}^\dagger(r_d)$ such that
\begin{equation}
	S_{m}^\mathrm{ab}=\frac{ e^{\alpha^\prime r_d^2 \exp\left(2r_s\right)}}{\cosh(r_s)}\left[H_{m}\left(-\frac{i\alpha^\prime r_d e^{r_s}}{2\sqrt{\beta}}\right)\right]^2\frac{\left(-\beta\right)^{m}}{m!},
\end{equation}
with $\alpha^\prime=\tanh r_s-1$.

We numerically illustrate the departure from such a statistics with various values of $\lambda_2$ in Figs.~\ref{fig2}(b)-(c). Given the commutator $\left[\mathcal{D}(r_s),\mathcal{S}(r_s)\right]\neq0$, the presence of the product $\mathcal{D}(r_s)\mathcal{S}(r_s)$ renders an asymmetry between the emission event $\ket{e;0_e}\rightarrow\ket{g;m_g}$ and the absorption event $\ket{g;0_g}\rightarrow\ket{e;m_e}$. Also, as a simple check, in the limiting case where $\lambda_{2}=0$, i.e., $\nu_{e}=\nu_{g}$, both transition strengths follow the same Poissonian distribution
$e^{-\lambda_{1}^2}\lambda_{1}^{2m}/m!$, as expected, reproducing the mirroring effect of emission and absorption spectra usually exhibited by most molecular transitions.
	
\section{Absorption and emission spectra}
	\label{QLE}

In order to derive spectroscopic quantities, we will assume a continuous wave classical drive coupled to the electronic transition incorporated in the following Hamiltonian
	\begin{equation}
		\mathcal{H}_\ell=i\eta_\ell\left(\sigma^{\dagger}e^{-i\omega_\ell t}-\sigma e^{i\omega_\ell t}\right),
	\end{equation}
with the Rabi frequency $\eta_\ell$ and laser frequency $\omega_\ell$. Since the molecule is also coupled to the electromagnetic vacuum and additional vibrational relaxation baths, we will make use of open system dynamics methods, first formulated in terms of a master equation. First, we include a spontaneous emission channel with the collapse operator $\sigma$ at rate $\gamma$. In addition, as the electronic transition is modified by the vibrational mode~\cite{mustroph2009relationship,miao2023deciphering,zirkelbach2022high}, the influence of the environment onto the dynamics of the vibrational mode can be well described by a collapse operator $\mathcal{U}b\mathcal{U}^\dagger$ at the rate $\Gamma$. For numerical investigations, the master equation for the system is given
\begin{equation} \dot{\rho}=-i\left[\mathcal{H}+\mathcal{H}_{\ell},\rho\right]+\mathcal{L}_{\gamma}[\sigma]\rho+\mathcal{L}_{\Gamma}\left[\mathcal{U}b\mathcal{U}^\dagger\right]\rho,\label{eq:master}
\end{equation}
where the standard Lindblad superoperator is written as $\mathcal{L}_{\gamma_\mathcal{O}}\cdot=\gamma_{\mathcal{O}}\left(2\mathcal{O}\cdot\mathcal{O}^{\dagger}-\mathcal{O}^{\dagger}\mathcal{O}\cdot-\cdot\mathcal{O}^{\dagger}\mathcal{O}\right)$ for a collapse operator $\mathcal{O}$ and a corresponding decay rate $\gamma_{\mathcal{O}}$. In particular in the polaron transformation $\tilde{\rho}=\mathcal{U}^{\dagger}\rho\mathcal{U}$, the last term in Eq.~(\ref{eq:master}) is going to the familiar form $\mathcal{L}_{\Gamma}\left[b\right]\tilde{\rho}$. The dot stands for the position where the density operator, on which the Lindblad superoperator is applied on, is to be included.

It is convenient, for deriving analytical results, to map the master equation into an equivalent set of QLEs. For any system operator $\mathcal{A}$ this can be done as follows~\cite{reitz2019langevin,gardiner2004quantum}
\begin{align}\nonumber
		\dot{\mathcal{A}}= & -i\left[\mathcal{A},\mathcal{H}+\mathcal{H}_{\ell}\right]-\left[\mathcal{A},\mathcal{O}^{\dagger}\right]\left(\gamma_{\mathcal{O}}\mathcal{O}-\sqrt{2\gamma_{\mathcal{O}}}\mathcal{O}_{\mathrm{in}}^{\phantom{\dagger}}\right)\\
		&+\left(\gamma_{\mathcal{O}}\mathcal{O}^{\dagger}-\sqrt{2\gamma_{\mathcal{O}}}\mathcal{O}_{\mathrm{in}}^{\dagger}\right)\left[\mathcal{A},\mathcal{O}\right],
	\end{align}
where $\mathcal{O}_{\mathrm{in}}$ is the zero-averaged and delta-correlated input noise operator associated with the collapse operator $\mathcal{O}$ and $\gamma_{\mathcal{O}}$ is the associated decay rate.

For molecules in solid-state environments or in solvents, the vibrational relaxation rate is usually very large greatly surpassing both $\gamma$ and $\eta_\ell$. Therefore, fluorescence occurs preferentially from the state $\ket{e,0_e}$, which lies at the bottom of the excited state manifold: this is generally referred to as Kasha's rule~\cite{del_Valle2019kasharule}. The same mechanism is valid for the absorption process, where absorption occurs from the state  $\ket{g,0_g}$, the lowest in energy. We will make use of this fast vibrational relaxation to impose a quick timescale for the modification of the bosonic $b$ operators and use their quasi-steady state values in the following. First, however, let us partition the total Hilbert space into two orthogonal subspaces (ground and excited electronic state manifolds) via the following two projection operators $\mathcal{P}_g=\sigma\sigma^\dagger$ and $\mathcal{P}_e=\sigma^\dagger\sigma$.
Let us first pay attention to the dynamical equation in the manifold of $\mathcal{P}_e$. For convenience reasons, we introduce a projected bosonic operator $b_e=\mathcal{U}b\mathcal{U}^\dagger\mathcal{P}_e$  acting only in this manifold and more explicitly expressed as
\begin{equation}
	b_e=\left(\cosh r_s b+\sinh r_s b^\dagger-r_de^{r_s}\right)\mathcal{P}_e
\end{equation}
and obeying the relation $b_e^\dagger b_e\ket{e;m_e}=m_e\ket{e;m_e}$. Meanwhile, we define a time-dependent generalized polaron operator~\cite{reitz2019langevin,reitz2020molecule}, by the transformation $\tilde{\sigma}_e=\sigma{\mathcal{S}}_e^\dagger{\mathcal{D}}_e^\dagger\exp\left[i(\nu_e-\nu_g)b^\dagger_e b_e^{\phantom{\dagger}} t\right]$. This allows the derivation of a set of effective QLEs in the rotating frame at the driving frequency $\omega_\ell$ for the emission process (see Appendix~\ref{B} for details)
\begin{subequations}
	\begin{align}
		&\dot{b}_e\approx-\left(i\nu_e+\Gamma\right)b_e+\sqrt{2\Gamma}\mathcal{B}^\mathrm{in}_e\mathcal{P}_e,\\
		&\begin{aligned}
			\dot{\tilde{\sigma}}_e\approx& -\left(i\Delta_\ell+\gamma\right)\tilde{\sigma}_e-\eta_\ell{\mathcal{S}}_e^\dagger{\mathcal{D}}_e^\dagger e^{i\left(\nu_e-\nu_g\right)b^\dagger_e b_e^{\phantom{\dagger}} t}\\
			&+\sqrt{2\gamma}\sigma_\mathrm{in}{\mathcal{S}}_e^\dagger{\mathcal{D}}_e^\dagger e^{i\left(\nu_e-\nu_g\right)b^\dagger_e b_e^{\phantom{\dagger}} t},\label{EQ:Polaron_E}
		\end{aligned}\\ &\dot{\mathcal{P}}_e=-2\gamma\mathcal{P}_e+\eta_\ell(\sigma+\sigma^\dagger)+\sqrt{2\gamma}(\sigma^\dagger\sigma_\mathrm{in}^{\phantom{\dagger}}+\sigma_\mathrm{in}^\dagger\sigma),\label{eq:Pop}
	\end{align}
\end{subequations}
with the detuning $\Delta_\ell=\omega_{00}-\omega_\ell$,  the displacement operator $\mathcal{D}_e=\exp\left[r_d(b_e^\dagger-b_e)\right]\mathcal{P}_e$ and the squeezing operator $\mathcal{S}_e=\exp\left[r_s(b_e^2-b_e^{\dagger2})/2\right]\mathcal{P}_e$. The input noises $\mathcal{B}_e^\mathrm{in}$ and $\sigma_\mathrm{in}$ are zero-averaged and have the following two-time correlations $\braket{\mathcal{B}_e^\mathrm{in}(t)\mathcal{B}_e^{\mathrm{in}\dagger}(t^\prime)}=\delta(t-t^\prime)$ and $\braket{\sigma_\mathrm{in}^{\phantom{\dagger}}(t)\sigma^\dagger_\mathrm{in}(t^\prime)}=\delta(t-t^\prime)$.

In a completely similar fashion, projected operators in the ground electronic state manifold can be defined. Let us introduce the ground state polaron operator via the transformation $\tilde{\sigma}_g=\exp{\left[i\left(\nu_e-\nu_g\right)b^\dagger_g b_g^{\phantom{\dagger}} t\right]}\mathcal{S}_g^\dagger\mathcal{D}_g^\dagger\sigma$ which allows one to derive a similar set of QLEs
\begin{subequations}
	\begin{align}
		& \dot{b}_g\approx-(i\nu_g+\Gamma)b_g+\sqrt{2\Gamma}\mathcal{B}_g^\mathrm{in}\mathcal{P}_g,\\
		& \begin{aligned}
			\dot{\tilde{\sigma}}_g\approx& -(i\Delta_\ell+\gamma)\tilde{\sigma}_g+\eta_\ell e^{i\left(\nu_e-\nu_g\right)b^\dagger_g b_g^{\phantom{\dagger}} t}\mathcal{S}_g^\dagger\mathcal{D}_g^\dagger\\
			&+\sqrt{2\gamma}e^{i\left(\nu_e-\nu_g\right)b^\dagger_g b_g^{\phantom{\dagger}} t}\mathcal{S}_g^\dagger\mathcal{D}_g^\dagger\sigma_\mathrm{in}.
			\label{EQ:Polaron_G}
		\end{aligned}
	\end{align}
\end{subequations}
As above, the new displacement operator is $\mathcal{D}_g=\exp[r_d(b^\dagger_g-b_g)]\mathcal{P}_g$, and the new squeezing operator is $\mathcal{S}_g=\exp[r_s(b^2_g-b^{\dagger2}_g)/2]\mathcal{P}_g$. The nonvanishing correlation of the zero-average noise operator is given by $\braket{\mathcal{B}_g^\mathrm{in}(t)\mathcal{B}_g^{\mathrm{in}\dagger}(t^\prime)}=\delta(t-t^\prime)$.

We are now in the position of reconstructing the full solution of the coherence operator in steady state by summing over the contributions in the ground and excited state manifolds. This can be done by formal integration of Eq.~(\ref{EQ:Polaron_E}) and Eq.~(\ref{EQ:Polaron_G}) to obtain a solution for $\braket{\sigma}$ expressed as
\begin{widetext}
	\begin{equation}
		\begin{aligned}
			\braket{\sigma}=&-\eta_\ell\int_0^\infty d\tau\, \Theta(t-\tau) e^{-(i\Delta_\ell+\gamma)\left(t-\tau\right)}\braket{\mathcal{S}_{e}^{\dagger}\left(\tau\right)\mathcal{D}_{e}^{\dagger}\left(\tau\right)e^{i\left(\nu_e-\nu_g\right)b_e^{\dagger}b_e^{\phantom{\dagger}} \tau}e^{-i\left(\nu_e-\nu_g\right)b_e^{\dagger}b_e^{\phantom{\dagger}} t}\mathcal{D}_{e}\left(t\right)\mathcal{S}_{e}\left(t\right)}\\
			&+\eta_\ell \int_0^\infty d\tau \, \Theta(t-\tau) e^{-(i\Delta_\ell+\gamma)\left(t-\tau\right)}\braket{\mathcal{D}_g(t)\mathcal{S}_g(t)e^{-i\left(\nu_e-\nu_g\right)b_g^\dagger b_g^{\phantom{\dagger}} t}e^{i\left(\nu_e-\nu_g\right)b_g^\dagger b_g^{\phantom{\dagger}} \tau}\mathcal{S}_g^\dagger(\tau))\mathcal{D}^\dagger_g(\tau)}.
		\end{aligned}
	\label{eq:Dipole_Formal}
	\end{equation}
Here, we have used the Heaviside step function $\Theta(t)$ and the initial value $\braket{\sigma(0)}=0$. Considering that the vibrational mode has a large relaxation rate (i.e., $\Gamma\gg\gamma$), we then decouple the vibronic and electronic degrees of freedom. The two-time correlation functions on the right side of above equation could be expressed as (see Appendix~\ref{B} for details)
\begin{subequations}
	\begin{align}
			\braket{\mathcal{D}_e(\tau)\mathcal{S}_e(\tau) e^{i(\nu_e-\nu_g)b^\dagger_e b_e^{\phantom{\dagger}} \tau} e^{-i(\nu_e-\nu_g)b^\dagger_e b_e^{\phantom{\dagger}} t}\mathcal{S}_e^\dagger(t)\mathcal{D}_e^\dagger(t)}=&\sum_{m=0}^{\infty}S_m^\mathrm{em} e^{-m(i\nu_g+\Gamma)(t-\tau)}\braket{\mathcal{P}_e\left(\tau\right)},\\
			\braket{\mathcal{D}_g(t)\mathcal{S}_g(t)e^{-i\left(\nu_e-\nu_g\right)b_g^\dagger b_g^{\phantom{\dagger}} t}e^{i\left(\nu_e-\nu_g\right)b_g^\dagger b_g^{\phantom{\dagger}} \tau}\mathcal{S}_g^\dagger(\tau))\mathcal{D}^\dagger_g(\tau)}=& \sum_{m=0}^{\infty}S_m^\mathrm{ab} e^{m(-i\nu_e+\Gamma)(t-\tau)}\braket{\mathcal{P}_g(\tau)}.
	\end{align}
	\end{subequations}
\end{widetext}
%%%%%%%%%%%%%%%%%%%%%%%
%%%%%%%%%%%%%%%%%%%%%%
\begin{figure}[t]
	\centering
	\includegraphics[width=0.95\columnwidth]{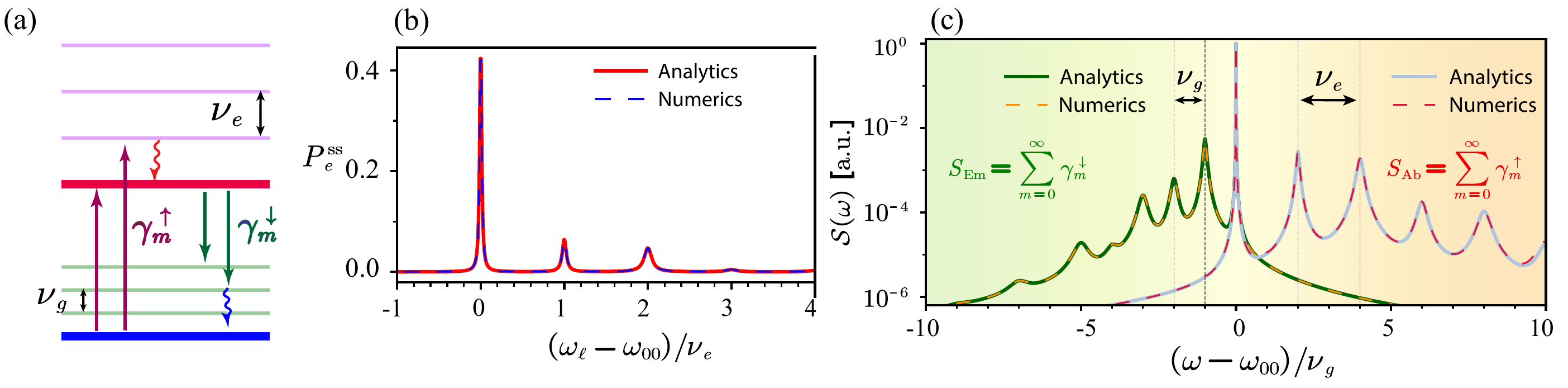}
	\caption{(a) Jablonski diagram illustrating possible emission and absorption processes. (b) Comparison of analytical and numerical results of the excited state population as a function of normalized detuning. Parameters are $\lambda_{2}=1$ (i.e., $\nu_e/\nu_g=2$), $\Gamma/\nu_g=0.1$, $\gamma/\Gamma=0.1$ and $\eta_{\ell}/\gamma=2$. (c) Comparison of analytical results versus numerical simulations for the absorption (shaded in orange) and emission (shaded in green) profiles. Parameters are $\eta_{\ell}/\gamma=0.1$ and $\Delta_\ell/\nu_{g}=0.1$.}
	\label{fig3}
\end{figure}
%%%%%%%%%%%%%%%%%%%%%%%%%
%%%%%%%%%%%%%%%%%%%%%%%%%
Replacing the infinite sums from above back into Eq.~(\ref{eq:Dipole_Formal}) leads to a convolution in time. This can be dealt with by employing a Laplace transformation defined as $\xoverline{f}(s)=\int_0^\infty dt\, f(t)\exp(-st)$ for a time-dependent function $f(t)$ at $t\geq0$. In such a case,
Eq.~(\ref{eq:Dipole_Formal}) takes a much simpler form
 \begin{equation}
\xoverline{\braket{\sigma}}=\frac{\eta_{\ell}}{s}\xoverline{\mathcal{G}}_\mathrm{ab}-\eta_{\ell}\xoverline{\braket{\mathcal{P}_e}}(\xoverline{{\mathcal{G}}}_\mathrm{em}+\xoverline{{\mathcal{G}}}_\mathrm{ab}),
 \end{equation}
with the following functions identified corresponding to emission and absorption events, respectively	
\begin{align}
	\xoverline{\mathcal{G}}_\mathrm{em}=&\sum_{m=0}^{\infty}\frac{S_{m}^\mathrm{em}}{s+m\Gamma+\gamma+i\left(\Delta_\ell-m\nu_g\right)},\\
	\xoverline{\mathcal{G}}_\mathrm{ab}=& \sum_{m=0}^{\infty}\frac{S_{m}^\mathrm{ab}}{s+\gamma+m\Gamma+i\left(\Delta_\ell+m\nu_e\right)}.
\end{align}
From these expressions, one can proceed in evaluating analytically the population of the excited state $p_e^\mathrm{ss}=\lim\limits_{t\to \infty}\braket{\sigma^\dagger(t)\sigma(t)}$ in steady state (as detailed in Appendix~\ref{App:SteadyPop})
 \begin{equation}
 	p_e^{\mathrm{ss}}=\frac{ \sum\limits_{m=0}^{\infty}\gamma_m^{\uparrow}(\omega_\ell)}{\gamma+\sum\limits_{m=0}^{\infty}\left[\gamma_m^{\uparrow}(\omega_\ell)+\gamma_m^{\downarrow}(\omega_\ell)\right]},\label{eq:PopE}
 \end{equation}
The coefficients $\gamma_m^{\uparrow}(\omega)$ and $\gamma_m^{\downarrow}(\omega)$ represent the dynamic equilibrium population transfer rates for absorption from the ground state to the excited state $\ket{g;0_g}\rightarrow\ket{e;m_e}$ and emission from the excited to the ground state $\ket{e;0_e}\rightarrow\ket{g;m_g}$ as illustrated in Fig.~\ref{fig3}(a).
%where the scaling of the vibrational rates has been on purpose exaggerated in order to clearly point out the difference in energies expected for the smaller and higher energy sidebands. This indicates a deviation from the standard physics of mirror symmetry in absorption and emission profiles.
The rates are analytically expressed as
 \begin{subequations}
 \begin{align}	\gamma^{\uparrow}_m(\omega)=&\frac{\eta_{\ell}^2S_m^\mathrm{ab}\left(m\Gamma+\gamma\right)}{(m\Gamma+\gamma)^2+(\omega_{0 0}+m\nu_e-\omega)^2},\\
		\gamma^{\downarrow}_m(\omega)=& \frac{\eta_{\ell}^2S_m^{\mathrm{em}}\left(m\Gamma+\gamma\right)}{(m\Gamma+\gamma)^2+(\omega_{00}-m\nu_g-\omega)^2}.
	\end{align}
 \end{subequations}
Specifically, these rates contribute to the rate equation for the population of the excited state, given by (see Appendix~\ref{App:RateEquation} for detailed derivations):
\begin{equation}
	\partial_t p_e=-2(\gamma+\sum\limits_{m=0}^{\infty}\gamma_m^\downarrow)p_e+2\sum_{m=0}^{\infty}\gamma_m^{\uparrow}(1-p_e).
\end{equation}
This equation holds true under the condition $\eta_\ell\ll\Gamma$. Remarkably, one could also obtain the same expression for the population of the excited state in steady state and compare with full numerical simulations to a very good fit, as illustrated in Fig.~\ref{fig3}(b). The parameters are given in the caption and are chosen in close attention to other works~\cite{chenu2019two,banerjee2002spectra}.

Additionally, we can employ the pump-probe scenario to analyze the absorption and emission processes. In this scenario, the molecule absorbs a photon at the frequency $\omega_\ell$, transitioning to the excited state $\ket{e;m_e}$ under the resonant condition $\omega_\ell=\omega_{00}+m\nu_e$. Subsequently, after undergoing fast vibrational relaxation, the molecule emits a photon centered around the frequency $\omega_{00}-m^\prime\nu_g$, which can be detected with a modified linewidth $\gamma+m^\prime\Gamma$. The absorption and emission profiles are then obtained by summing up the contributions from all possible cases, resulting in Lorentzian profiles represented by $\gamma_m^{\uparrow/\downarrow}$, as shown in Fig.~\ref{fig3}(c):
\begin{equation}
	S_{\mathrm{Ab}}=\sum_{m=0}^\infty \gamma^\uparrow_{m}\quad \mathrm{and} \quad	S_{\mathrm{Em}}=\sum_{m=0}^\infty \gamma^\downarrow_{m}.
\end{equation}
Here, the scaling of the vibrational rates has been on purpose exaggerated in order to clearly point out the difference in energies expected for the smaller and higher energy sidebands. The presence of the quadratic electron-vibron coupling under realistic conditions, is expected to only slightly break the symmetry between the emission and absorption spectra, as the expected values for $\lambda_2$ lie well below in the subunit region. More details on the procedure we have followed for the above derivations is presented in Appendix~\ref{App:Ab&Em} and basically follows the quantum regression theorem formalism~\cite{gardiner2004quantum,carmichael1999statistical}.
%%%%%%%%%%%%%%%%%%%%%
%%%%%%%%%%%%%%%%%%%%
\begin{figure}[b]
	\includegraphics[width=0.60\columnwidth]{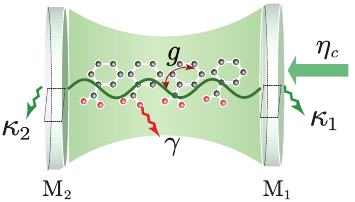}
	\caption{Schematics of an ensemble of molecules inside a Fabry-P\'{e}rot resonator. Cavity photon loss occurs at rates $\kappa_1$ and $\kappa_2$ via the mirrors M1 and M2, respectively. Light-molecule interactions occur at rate $g$ while spontaneous emission and cavity driving are at rates $\gamma$ and $\eta_c$, respectively.}
	\label{fig4}
\end{figure}
%%%%%%%%%%%%%%%%%%%%%
%%%%%%%%%%%%%%%%%%%%
\section{Molecular Polaritonics}
\label{sec:Molecular-Cavity}
Let us now ask what is the imprint of the asymmetry between the ground and excited state potential landscapes on the signal of an optical cavity containing such a molecule in the strong coupling regime of cavity quantum electrodynamics. To this end, we consider a single molecule placed within the optical volume of a single mode optical cavity mediating transitions between the ground and excited potential landscapes. Under strong optical confinement conditions, the interaction of light and matter can lead to the production of hybrid quantum states, i.e., polaritons~\cite{wu2016when,pino2018tensor,delpino2019tensor,keeling2018organic,sentef2018cavity,herrera2020molecular,neuman2018origin,zeb2018exact,kansanen2019theory,du2018theory,ribeiro2018polariton} as superpositions of ground or excited electronic states and zero or single photon states. While polaritons are eigenstates solely of the electron-photon interaction Hamiltonian, the intrinsic electron-vibron coupling can provide a mechanism of polariton cross-talk, leading to a unidirectional loss of energy from the higher state to the lower energy state. This has been shown analytically in Ref.~\cite{reitz2019langevin} for the standard case of identical ground and excited state potential landscapes and found to be most pronounced when the vibrational mode is resonant to the interpolariton frequency splitting.

Let us now consider the case of $\mathcal{N}$ molecules inside the spatial extent of a single-mode of a Fabry-P\'{e}rot optical resonatoras, illustrated in Fig.~\ref{fig4}. The dynamics of a single molecule is governed by the Hamiltonian $\mathcal{H}$ from  Eq.~\eqref{eq:Ham}. The interaction between the $\mathcal{N}$ molecules and the cavity field mode is characterized by the Tavis-Cummings model,
\begin{equation}
	\begin{aligned}
			\mathcal{H}_\mathrm{cav}=&\omega_{c} a^\dagger a+g\sum_{n=1}^\mathcal{N}\left(a\sigma^\dagger_n +\mathrm{h.c.}\right)\\
			&+i\eta_{c} (a^\dagger e^{-i\omega_\ell t}-\mathrm{h.c.}),\label{eq:Ham_TC}
	\end{aligned}
\end{equation}
consisting of the free cavity field at frequency $\omega_c$ and with bosonic mode $a$ and the Tavis-Cummings interaction with the unit light-matter coupling strength $g$ and the laser field drive with amplitude
$\eta_c$ and frequency $\omega_\ell$.For convenience, we have made the assumption here that all molecules are identical. Let us proceed with a set of effective QLEs for the cavity mode $a$ and the state dependent polaron operators $\tilde{\sigma}_{e,n}$ and $\tilde{\sigma}_{g,n}$ for the $n$th molecule in the rotating frame at the laser frequency $\omega_\ell$:
 \begin{subequations}
 	\begin{align}
 		&\dot{a}=-(i\Delta_c+\kappa)a-ig\sum_{n=1}^\mathcal{N}\sigma_n+\sqrt{2\kappa_1}A_\mathrm{1,in}+\sqrt{2\kappa_2}a_\mathrm{2,in},\label{eq:Cav}\\
 		&\begin{aligned}
 			\dot{\tilde{\sigma}}_{e,n}\approx& -\left(i\Delta_\ell+\gamma\right)\tilde{\sigma}_{e,n}+iga{\mathcal{S}}_{e,n}^\dagger{\mathcal{D}}_{e,n}^\dagger e^{i\left(\nu_e-\nu_g\right)b^\dagger_{e,n} b_{e,n} t}\\
 			&+\sqrt{2\gamma}\sigma_\mathrm{in,n}{\mathcal{S}}_{e,n}^\dagger{\mathcal{D}}_{e,n}^\dagger e^{i\left(\nu_e-\nu_g\right)b^\dagger_{e,n} b_{e,n} t},\label{eq:Polaron_E}
 		\end{aligned}\\
 		& \begin{aligned}
 			\dot{\tilde{\sigma}}_{g,n}\approx& -(i\Delta_\ell+\gamma)\tilde{\sigma}_{g,n}-iga e^{i\left(\nu_e-\nu_g\right)b^\dagger_{g,n} b_{g,n} t}\mathcal{S}_{g,n}^\dagger\mathcal{D}_{g,n}^\dagger\\
 			&+\sqrt{2\gamma}e^{i\left(\nu_e-\nu_g\right)b^\dagger_{g,n} b_{g,n} t}\mathcal{S}_{g,n}^\dagger\mathcal{D}_{g,n}^\dagger\sigma_\mathrm{in,n}^{\phantom{\dagger}}.
 			\label{eq:Polaron_G}
 		\end{aligned}
 	\end{align}
 \end{subequations}
Here, the total dissipation for the cavity field $\kappa=\kappa_1+\kappa_2$ encompasses the losses via both mirrors. The operators $A_\mathrm{1,in}=\eta_{c}/\sqrt{2\kappa_1}+a_\mathrm{1,in}$ describes the input classical field coming through the left mirror $\eta_{c}/\sqrt{2\kappa_1}$ and the zero-average input noise with the only non-vanishing two-time correlations
$\braket{a_\mathrm{1,in}(t)a_\mathrm{1,in}^\dagger(t^{\prime})}=\delta(t-t^\prime)$. Additionally, zero-average input noise comes through the right side mirror with similar correlations
$\braket{a^{\phantom{\dagger}}_\mathrm{2,in}(t)a_\mathrm{2,in}^\dagger(t^{\prime})}=\delta(t-t^\prime)$ and uncorrelated with the
$a_\mathrm{1,in}(t)$.\\

The Markovian limit is achieved under the large relaxation rate condition for vibrational mode, i.e., $\Gamma\gg\kappa$ and $\Gamma\gg\gamma$. In this case, the approach to treat the vibrations as a local phonon bath is still applicable. By formally integrating the equations for polaron operator, tracing over the cavity mode as well as electronic degrees of freedom and taking the Laplace transformation, we have
\begin{equation}
	\begin{aligned}
		\xoverline{\left\langle \sigma_n\right\rangle} =&ig\left(\xoverline{\mathcal{G}}_{\mathrm{em}}+\xoverline{\mathcal{G}}_{\mathrm{ab}}\right)\xoverline{\left\langle\mathcal{P}_{e,n}a\right\rangle} -ig\xoverline{\mathcal{G}}_{\mathrm{ab}}\xoverline{\left\langle a\right\rangle}.
	\end{aligned}\label{Eq:Formal_Pauli_Cavity}
\end{equation}
%%%%%%%%%%%%%%%%%%%%%%%
%%%%%%%%%%%%%%%%%%%%%%%
\begin{figure}[t]
	\includegraphics[width=0.95\columnwidth]{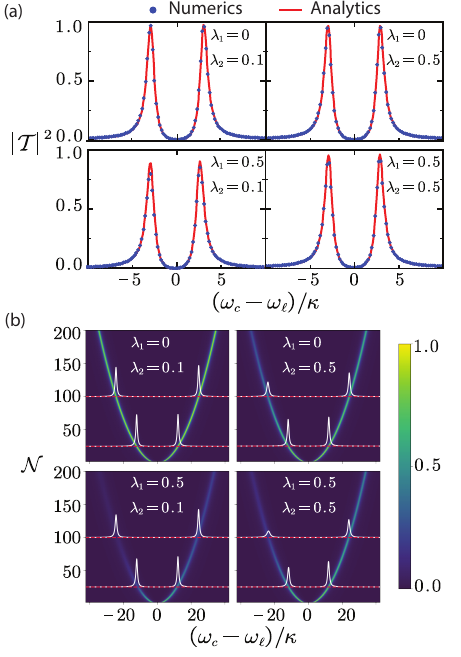}
	\caption{Cavity transmission ($|\mathcal{T}|^2$) of the molecule with various  linear and quadratic electron-vibron couplings $\lambda_1$ and $\lambda_2$ for a strong coupling to a single cavity mode. (a) Cavity transmission with a single molecule. (b) Cavity transmission as the function of the number of molecules in the cavity. The white lines represent the profile at $\mathcal{N}=25$ and $100$, respectively. Parameters: $\omega_{c}=\omega_{00},\,g=3\kappa,\,\nu_g=10\kappa,\,\gamma=0.01\kappa, \Gamma=20\kappa,\,2\kappa_1=2\kappa_2=\kappa$, and the driving is assumed very weak $\eta_{c}/\kappa=0.001$.}
	\label{fig5}
\end{figure}
%%%%%%%%%%%%%%%%%%%%%
%%%%%%%%%%%%%%%%%%%%
The coupling between the cavity mode $a$ and the projection operator $\mathcal{P}_{e,n}$ leads to non-linear effects. However, we restrict our analysis to the weak excitation regime, i.e., the cavity photon number is much smaller than unity and the population of the excited electronic state $\ket{e}$ is negligible (under the condition that $\eta_c\ll\kappa$). In other words, this approximation allows for the construction of a linear response theory formalism where the transmitted light gives information on the position and linewidths of the hybrid light-matter eigenstates of the system. In the case of identical conditions, the expectation value of the electronic coherence operator $\sigma_n$ for the $n$-th molecule will be equivalent to that of the other molecules, i.e., $\braket{\sigma}=\braket{\sigma_n}=\braket{\sigma_m}$ ($m\neq n$). Then, the equations of motion are written in the vector form (in the Laplace transform domain) as  $\xoverline{\mathbf{M}}\xoverline{\mathbf{v}}+\xoverline{\mathbf{v}}_{c}=0$, with the drift matrix
	\begin{equation}
		\xoverline{\mathbf{M}}=\left(\begin{array}{ccc}
			-\left(i\Delta_{c}+\kappa\right)-s & -i\mathcal{N}g \\
			-ig & -1/ \overline{\mathcal{G}}_{\mathrm{ab}}
		\end{array}\right),
	\end{equation}
and the definitions $\mathbf{v}=\left(\braket{a},\braket{\sigma}\right)^T$ and $\mathbf{v}_c=\left(\eta_{c},0\right)^T$.
The diagonalization of the drift matrix (under resonance condition $\Delta_{c}=0$) yields the frequencies $\omega_\pm$ and linewidths $\gamma_\pm$~\cite{plankensteiner2019enhanced,herrera2020molecular} of the two polaritons as
\begin{subequations}		
	\begin{align}		
    \omega_\pm=&\frac{-\Delta_\mathrm{eff}}{2}\pm\frac{1}{2}\mathcal{I}\sqrt{(\Gamma_\mathrm{eff}-\kappa+i\Delta_\mathrm{eff})^2-\mathcal{N}g^2},\\
	\gamma_\pm=&\frac{\Gamma_\mathrm{eff}+\kappa}{2}\pm\frac{1}{2}\mathcal{R}\sqrt{(\Gamma_\mathrm{eff}-\kappa+i\Delta_\mathrm{eff})^2-\mathcal{N}g^2},
	\end{align}
	\end{subequations}	
with $\Gamma_\mathrm{eff}=\mathcal{R}\lim\limits_{s\rightarrow0}1/\overline{\mathcal{G}}_{\mathrm{ab}}$ and  $\Delta_\mathrm{eff}=\mathcal{I}\lim\limits_{s\rightarrow0}1/\overline{\mathcal{G}}_{\mathrm{ab}}$ denoting the effective decay rate and additional frequency shift. These particularities of the polaritons can be explored in a very simple way by performing a scan of the laser frequency around the cavity resonance and noticing the position of the peaks corresponding to the hybrid light-matter states. This can be done at the analytical level in the weak excitation regime and compared to full exact numerics. We define the complex cavity transmission amplitude as the ratio of the normalized continuous outgoing field versus incoming field amplitudes
\begin{equation}
	\mathcal{T}=\frac{\sqrt{2\kappa_2}\braket{a}_\mathrm{ss}}{\eta_{c}/\sqrt{2\kappa_1}},
\end{equation}
and illustrate its behavior with respect to the scanning laser frequency in Fig.~\ref{fig5}. The quantity $\braket{a}_\mathrm{ss}$ is the average value of the cavity mode amplitude in steady state in the linear response regime
	\begin{align}
		\braket{a}_\mathrm{ss}=& \frac{\eta_{c}}{\mathcal{N}g^2\chi_\mathrm{ab}+\kappa+i(\omega_{c}-\omega_\ell)},
	\end{align}
with $\chi_\mathrm{ab}=\lim\limits_{s\rightarrow0}1/\overline{\mathcal{G}}_{\mathrm{ab}}.$

We illustrate numerical and analytical results in Fig.~\ref{fig5} where the profile of the cavity transmission at $\omega_{c}=\omega_{00}$ is plotted. The presence of the linear electron-vibron coupling scaling with $\lambda_1$ induces an interaction between upper and lower polaritons already presented at the theoretical level in a few treatments~\cite{neuman2018origin,reitz2019langevin,reitz2020molecule}. Instead, at the level of a single molecule, the quadratic interaction will suppress the polariton cross talk, as illustrated in Fig~\ref{fig5}(a). In essence, the squeezing term is responsible with a shift in the molecular resonance which then in turn brings the cavity off-resonance with the electronic transition except the zero-phonon transition process. Increasing the number of molecules while assuming very weak driving conditions presents a different situation. This is shown in Fig.~\ref{fig5}(b) as an effective reduction of the upper polariton with increasing particle number.\\

\section{Conclusions}
We have applied the toolbox of open system dynamics and in particular the QLEs formalism to analytically describe spectroscopic properties of solid-state embedded molecules in free space or in optical cavity settings. In particular, we generalized our previous approach introduced in Ref.~\cite{reitz2019langevin} to a scenario where the potential landscapes of a molecule have unequal curvatures in the ground and excited electronic state. This has seen the introduction of a generalized polaron operator where the electronic degree of freedom is dressed by vibrations via a displacement operation followed by an additional squeezing operation. The first effect is seen in the emergent asymmetry between absorption and emission profiles for molecular spectroscopy. A second effect that emerges from our analytical calculations is the context of cavity quantum electrodynamics where the additional squeezing operation leads to a detuning between the bare molecular resonance and the cavity resonance. Our calculations can be relevant in the direction of optomechanics or optovibronics, owing to the strong electron-vibron couplings, albeit under very lossy conditions.
	
\section*{Acknowledgments }
We acknowledge financial support from the Max Planck Society and from the Deutsche Forschungsgemein-
schaft (DFG, German Research Foundation) – Project-ID 9429529648 – TRR 306 QuCoLiMa (”Quantum Cooperativity of Light and Matter”).

	\bibliography{Refs}
	
	\newpage
	\onecolumngrid
	\newcommand{\non}{\nonumber}
	
	\newpage
	\appendix

\section{The modified Holstein Hamiltonian}
	Let us illustrate how the modified Holstein Hamiltonian arises, how it can be diagonalized and how the quantum Langevin equations for the squeezed and displaced polaron operators can be derived.
%%%%%%%%%%%%%%%%%%%%%%%%%%%%%%%%%%%%%%%%%%%%%%%%%%%%%%%%%%%%%%%%%%%%%%
\subsection*{First principle derivation of the non-linear Hamiltonian}
%%%%%%%%%%%%%%%%%%%%%%%%%%%%%%%%%%%%%%%%%%%%%%%%%%%%%%%%%%%%%%%%%%%%%%
	We consider a single molecule with ground $|g \rangle$ (frequency $\omega_g$) and excited $|e \rangle$ (frequency $\omega_e$) electronic levels coupled to the ground phonons ($\nu_g$) and the excited phonons ($\nu_e$)  of a single vibrational mode with mass $\mu$ respectively. Assuming that the ground and excited electronic states have different parabolic shape around the minima. Then the total Hamiltonian of the electron-phonon system reads as ($\hbar=1$).
	\begin{eqnarray}\nonumber
		\mathcal{H}&= &  \left[\omega_e+ \dfrac{{\hat P}^2}{2\mu}  +\dfrac{1}{2}\mu\nu^2_e (\hat R-R_\mathrm{eg})^2  \right] \sigma^\dagger  \sigma
		+\left(\omega_g+\dfrac{{\hat P}^2}{2\mu}  +\dfrac{1}{2}\mu\nu^2_g {\hat R}^2 \right) \sigma \sigma^\dagger \\
		&=& \omega_g   \sigma  \sigma^\dagger + \left(\omega_e + \mu\nu^2_e R_\mathrm{eg}^2/2 \right)\sigma^\dagger  \sigma + \dfrac{{\hat P}^2}{2\mu}  +\dfrac{1}{2}\mu \nu^2_g {\hat R}^2+\dfrac{1}{2}\mu \nu^2_e R_{\mathrm{eg}} \hat R\sigma^\dagger  \sigma+  \dfrac{1}{2}\mu(\nu^2_e-\nu^2_g){\hat R}^2\sigma^\dagger  \sigma,
		\label{Hm}
	\end{eqnarray}
	where $\sigma= |g \rangle \langle e|$ is the Pauli lowering operator. By rewriting position ($\hat R$) and momentum ($\hat P$) in terms of creation $b^\dagger=(\hat R/{R_{\mathrm{zpm}}}-iR_{\mathrm{zpm}}\hat P)/\sqrt{2}$ and annihilation  $b=(\hat R/{R_{\mathrm{zpm}}}+iR_{\mathrm{zpm}}\hat P)/\sqrt{2}$  operators that fulfill $\left[b, b^\dagger\right]=1$, the Hamiltonian in Eq.~(\ref{Hm}) can be written as
	\begin{align}
		\mathcal{H}=\omega_{0}\sigma^\dagger  \sigma + \nu_g b^\dagger b   + \lambda_1\nu_g (b + b^\dagger ) \sigma^\dagger  \sigma  +  \lambda_2 \nu_g (b + b^\dagger )^2    \sigma^\dagger  \sigma,  \label{Hm1}
	\end{align}
	where ${\omega}_{0}=\omega_e-\omega_g + \mu\nu^2_e R_\text{eg}^2/2=\omega_e-\omega_g+\lambda_{1}^2\nu_g^3/\nu_e^2$ is the modified frequency of electronic excited state.  $\lambda_1  =   \mu\nu_{e}^{2}R_\mathrm{eg}R_{\mathrm{zpm}}/\nu_{g}$ and $\lambda_2 =  (\nu_e^2-\nu_g^2)/4\nu^2_g$ are the linear and quadratic coupling constants.
	
	%%%%%%%%%%%%%%%%%%%%%%%%%%%%%%%%%%%%%%%%%%%%%%%%%%%%%%%%%%
	%%%%%%%%%%%%%%%%%%%%%%%%%%%%%%%%%%%%%%%%%%%%%%%%%%%%%%%%%%
	%%%%%%%%%%%%%%%%%%%%%%%%%%%%%%%%%%%%%%%%%%%%%%%%%%%%%%%%%%
	
	\subsection*{Quadratic Holstein Hamiltonian Diagonalization}
	\label{sub:HamiltonianDiagonalize}
	In the presence of both linear and  quadratic couplings, the diagonalization of Hamiltonian in Eq.~(\ref{eq:Ham}) can be achieved by performing a sequence of unitary transformations. This transformation could be accomplished by first removing all the linear terms via the polaron transformation
		$\mathcal{U}_d={[\mathcal{D}(r_d)}]^{\sigma^\dagger \sigma}=\sigma \sigma^\dagger+\mathcal{D}(r_d)\sigma ^\dagger \sigma$,  where displacement operator is define as ${\mathcal{D}(r_d)}=\exp[r_d (b^\dagger-b)]$. The polaron transformation has the effect that $b\rightarrow \mathcal{U}_d^\dagger  b \mathcal{U}_d =b+r_d\sigma^\dagger \sigma$. Specifically, when
		\begin{equation}
			r_d=-\frac{\lambda_1}{1+4\lambda_2}=-\lambda_{1}\frac{\nu_g^2}{\nu_e^2},
		\end{equation}
	  the resulting Hamiltonian $	\mathcal{H}_1=\mathcal{U}_d^\dagger \mathcal{H}\mathcal{U}_d$ can be written as
	\begin{eqnarray}
		\mathcal{H}_1=&\left[{\omega}_0+r^2_d \nu_g (1+4 \lambda_2)  +2 \lambda_1 r_d  \right]\sigma^\dagger  \sigma + \nu_g b^\dagger b  +  \lambda_2 \nu_g  (b + b^\dagger)^2    \sigma^\dagger  \sigma.
	\end{eqnarray}
	
	We now diagonalize this Hamiltonian via squeezing transformation
$\mathcal{U}_s ={[\mathcal{S}(r_s)]}^{\sigma^\dagger \sigma}=\sigma \sigma^\dagger+\mathcal{S}(r_s)\sigma ^\dagger \sigma$, where ${\mathcal{S}(r_s)}=  \exp[r_s (b^2-b^{\dagger^2})/2]$ is a single mode squeezing operator, so as to remove the quadratic terms. Under this transformation $b\rightarrow \mathcal{U} ^\dagger_s b  \mathcal{U} _s=b \sigma \sigma^\dagger +\left[b \cosh(r_s) +b^\dagger \sinh(r_s)\right] \sigma^\dagger \sigma$, the resulting Hamiltonian $\mathcal{H}_2=\mathcal{U}^{\dagger}_s\mathcal{H}_1\mathcal{U}_s$ can be written as
	\begin{eqnarray}
		\mathcal{H}_2&=& \nu_g  b^\dagger b\sigma\sigma^\dagger+  \nu_e  b^\dagger b \sigma^\dagger \sigma  + {\omega}_{00} \sigma^\dagger \sigma
		\label{sq2}
	\end{eqnarray}
under the condition of
\begin{equation}
	e^{4r_s}=1+4\lambda_{2}=\frac{\nu_e^2}{\nu_g^2}.
\end{equation}
Here $\omega_{00}=\omega_{0} +r_d\lambda_1\nu_g+(\nu_e-\nu_g)/2$ is the zero-phonon line.
This transformation could be also accomplished in a reverse order, by first removing quadratic terms under applying $\mathcal{U}_s^\prime = [\mathcal{S}(r_s)]^{\sigma ^\dagger \sigma }$ and then removing the all linear terms via polaron transformation  $\mathcal{U}_d^\prime= [\mathcal{D}(r_d e^{r_s} )]^{\sigma ^\dagger \sigma }$. The results are, of course, identical.
	%%%%%%%%%%%%%%%%%%%%%%%%%%%%%%%%%%%%%%%%%%%%%%%%%%%%%%%%%%%%%%%%%%%%%
	\section{Derivation of the Effective Quantum Langevin Equations for the Electronic Transition}
	\label{B}
	
	 \subsection*{Effective Quantum Langevin Equation for the Vibrational Mode.}
	
	 	Let us consider one special case where the relaxation rate for the vibrations is much larger than the rate of change for the population on the electric excited state. Then the dynamical behaviors for vibrations with molecules populating on the state $\ket{e}$  prefer to being different with that with molecules occupying in the state $\ket{g}$. This fact motivates the partitioning of the total Hilbert space into the orthogonal subspaces via the following two projection operators
	 \begin{equation}
	 	\mathcal{P}_e=\sigma^\dagger\sigma
	 	\quad\text{and}\quad
	 	\mathcal{P}_g=\sigma\sigma^\dagger.
	 \end{equation}
	 The bosonic annihilation operator can thus be partitioned into
	 %\begin{equation}
	 $b=b\sigma^\dagger\sigma+b\sigma\sigma^\dagger$, which gives two dynamical equations  corresponding to the operator $b_1=b\sigma^\dagger\sigma$ and $b_2=b\sigma\sigma^\dagger$ as
	 	 \begin{subequations}
	 \begin{align}
	 	\dot{b}_1=&\,b\frac{d}{dt}(\sigma^{\dagger}\sigma)-\left[i\left(1+2\lambda_{2}\right)\nu_g+\Gamma\right]b_{1}-i\lambda_{1}\nu_g\sigma^{\dagger}\sigma-i2\lambda_{2}\nu_g b_{1}^{\dagger}+\Gamma r_{d}\mathcal{P}_e+\sqrt{2\Gamma}\mathcal{B}_{1}^{\mathrm{in}}\mathcal{P}_e,\label{eq:Vib_1}\\
	 	\dot{b}_2=&-b\frac{d}{dt}(\sigma^{\dagger}\sigma)-\left(i\nu+\Gamma\right)b_{2}+\sqrt{2\Gamma}\mathcal{B}_{2}^{\mathrm{in}}\mathcal{P}_g,\label{eq:Vib_G}
	 \end{align}
	 \end{subequations}
 where $\mathcal{B}_1^\mathrm{in}$ and  $\mathcal{B}_2^\mathrm{in}$ are the noise operators.

 	Assuming a large relaxation rate for the vibrational mode, the evolution of $\sigma^{\dagger}\sigma$  can be approximately negligible, while the vibrational mode rapidly relaxes to the steady situation. Then we can get the effective dynamical equations of $b_1$ and $b_2$
 	\begin{subequations}
 \begin{align}
 	\dot{b}_1&\approx-\left[i\left(1+2\lambda_{2}\right)\nu_g+\Gamma\right]b_{1}-i\lambda_{1}\nu_g\mathcal{P}_e-i2\lambda_{2}\nu_g b_{1}^{\dagger}+\Gamma r_{d}\mathcal{P}_e+\sqrt{2\Gamma}\mathcal{B}_{1}^{\mathrm{in}}\mathcal{P}_e,\label{eq:Vib_Eff_E}\\
 	\dot{b_2}&\approx-\left(i\nu+\Gamma\right)b_{2}+\sqrt{2\Gamma}\mathcal{B}_{2}^{\mathrm{in}}\mathcal{P}_g\label{eq:Vib_Eff_G}.
 \end{align}
 \end{subequations}
 By introducing $b_e=\mathcal{U} b_1\mathcal{U}^\dagger=\cosh r_s b_1+\sinh r_s b^\dagger_1-r_d\exp({r_s})\mathcal{P}_e$,
one can get the effective Langevin equation for $b_e$
\begin{equation}
	\frac{d}{dt}b_e=-(i\nu_e+\Gamma)b_e+\sqrt{2\Gamma}\mathcal{B}_\mathrm{e}^{\mathrm{in}}\mathcal{P}_e.\label{eq:Eff_BE}
\end{equation}
Under the condition $\Gamma\gg\gamma$, and $\Gamma\gg\eta_{\ell}$, one can assume the correlation between noise operators $\mathcal{B}_e^{\mathrm{in}{}}$ and $\mathcal{B}_g^{\mathrm{in}{}}$ is negligible as
	\begin{align}
		\braket{\mathcal{B}_e^{\mathrm{in}{}\dagger}(t)\mathcal{B}_g^{\mathrm{in}{}}(\tau)}\approx0\quad \mathrm{and}\quad \braket{\mathcal{B}_g^{\mathrm{in}{}\dagger}(t)\mathcal{B}_e^{\mathrm{in}{}}(\tau)}\approx0,
	\end{align}
and the nonvanishing correlation functions obeying the fluctuation-dissipation relation
\begin{equation}
	\braket{\mathcal{B}_e^{\mathrm{in}{}\dagger}(t)\mathcal{B}_e^{\mathrm{in}{}}(\tau)}=\delta(t-\tau)\quad \mathrm{and}\quad\braket{\mathcal{B}_g^{\mathrm{in}{}\dagger}(t)\mathcal{B}_g^{\mathrm{in}{}}(\tau)}=\delta(t-\tau).
\end{equation}
%Notably, Eq.~(\ref{eq:Emiss_Groups}) only consider the contribution of the vibrational mode in the manifold of $\mathcal{P}_e$. To obtain the information of the whole Hilbert space, we also need to get the dynamics in the manifold of $\mathcal{P}_g$ thorough the equations for $b_g=b\sigma\sigma^\dagger$ and $\tilde{\sigma}_g=\mathcal{S}_e^\dagger\mathcal{D}_e^\dagger\sigma$
%\begin{subequations}
%	\begin{align}
%			\dot{b_g}&\approx-\left(i\nu+\Gamma\right)b_{g}+\sqrt{2\Gamma^\prime}\mathcal{B}_{g}^{\mathrm{in}},\\
%			\dot{\sigma}_{g}&\approx - \left[i\Delta_g+i(\nu_e-\nu_g)b_g^\dagger b_g\gamma\right]\sigma_{g}+\eta_{\ell}\mathcal{S}^{\dagger}_{g}\mathcal{D}_{g}^{\dagger}+\sqrt{2\gamma}\mathcal{S}^{\dagger}_{g}\mathcal{D}_{g}^{\dagger}\sigma_{\mathrm{in}}\label{eq:Dot_Sigma_G}.
%		\end{align}
%\end{subequations}

%%%%%%%%%%%%%%%%%%%%%%%%%%%%%%%%%%%%%%%%%%%%%%%%%%%%%%%%%%%%%%%%%%%%

%%%%%%%%%%%%%%%%%%%%%%%%%%%%%%%%%%%%%%%%%%%%%%%%%%%%%%%%%%%%%%%%%%%%%

\subsection*{Effective Quantum Langevin Equation for the Electronic Transition.}
\label{sub:ET-EQLE}
%As introduced in Subsec.~(\ref{sub:HamiltonianDiagonalize}), the Quadratic Hamiltonian in Eq.~(\ref{Hm1}) performs the diagonalization form under the unitary transformation $\mathcal{U}$, while the dipole operator $\sigma$ is dressed via the vibrational mode as
%$\sigma\mathcal{DS}$.

Let us pay attention to the electronic transition by introducing the ``dressed" dipole operator, i.e., polaron operator, $\tilde{\sigma}_e^\prime=\sigma\mathcal{S}^\dagger_1\mathcal{D}^\dagger_1$. The Langevin equation  $\dot{\tilde{\sigma}}_e^\prime=\dot{\sigma}\mathcal{S}^\dagger_1\mathcal{D}^\dagger_1+\sigma\partial_t(\mathcal{S}^\dagger_1\mathcal{D}^\dagger_1)$ in a rotating frame at driving frequency $\omega_\ell$ can be expressed as
\begin{align}\nonumber
	\dot{\tilde{\sigma}}_e^\prime\approx&-\left[i(\omega_{00}-\omega_\ell)+\gamma\right]\tilde{\sigma}_e^\prime-\eta_{\ell}\mathcal{S}^\dagger_1\mathcal{D}^\dagger_1\sigma^\dagger\sigma+\sqrt{2\gamma}\sigma_\mathrm{in}\mathcal{S}^\dagger_1\mathcal{D}^\dagger_1\\
	&-i\left(1-\frac{\nu_g}{\nu_e}\right)\tilde{\sigma}_e^\prime\left[\nu_gb_1^\dagger b_1+\lambda_{1}\nu_g(b_1+b_1^\dagger)+\lambda_{2}\nu_g(b_1+b_1^\dagger)^2\right],\label{eq:QudraticTerm}
\end{align}
with $\mathcal{S}_1= \exp[r_s (b_1^2-b_1^{\dagger^2})/2]\mathcal{P}_e$ and $\mathcal{D}_{1}=\mathrm{exp}\left[r_d(b^\dagger_1-b_1)\right]\mathcal{P}_e$.

Considering that the quadratic terms in the second line of the equation above, i.e., $\nu_gb_1^\dagger b_1+\lambda_{1}\nu_g(b_1+b_1)+\lambda_{2}\nu_g(b_1+b_1^\dagger)^2$, can be reformed as $\nu_e b_e^{\dagger} b_e$, where the definition of $b^{}_e$ is introduced in Eq.~(\ref{eq:Eff_BE}), we can thus reform the equation above into
\begin{equation}
	\dot{\tilde{\tilde{\sigma}}}_e^{}\approx-\left[i(\omega_{00}-\omega_\ell)+\gamma\right]\tilde{\tilde{\sigma}}^{}_e-i(\nu_e-\nu_g)\tilde{\tilde{\sigma}}^{}_e  b^{{}\dagger}_eb_e-\eta_{\ell}\mathcal{S}^{{}\dagger}_e\mathcal{D}^{{}\dagger}_e\sigma^\dagger\sigma+\sqrt{2\gamma}\sigma_\mathrm{in}\mathcal{S}^{{}\dagger}_e\mathcal{D}^{{}\dagger}_e,\label{Eq:Polaron_E}
\end{equation}
with $\tilde{\tilde{\sigma}}_e^{}=\sigma\mathcal{S}^{{}\dagger}_e\mathcal{D}^{{}\dagger}_e$, $\mathcal{S}^{{}}_e=\exp\left[\left(r_s(b^{{}2}-b^{{}\dagger2})/2\right)\sigma^\dagger\sigma\right]\mathcal{P}_e$, and $\mathcal{D}^{{}}_e=\exp\left[r_d(b^{{}\dagger}-b^{})\sigma^\dagger\sigma\right]\mathcal{P}_e$.

  Taking account of its dynamical equation for  $\exp\left[i\left(\nu_e-\nu_g\right) b^{{}\dagger}_eb^{}_et\right]$ given by
\begin{align}\nonumber
	\frac{d}{dt}e^{i\left(\nu_e-\nu_g\right) b^{{}\dagger}_eb^{}_et}=& \,i\left(\nu_e-\nu_g\right)\int_{0}^{1}e^{i\alpha\left(\nu_e-\nu_g\right) b^{{}\dagger}_eb^{}_et}\frac{d}{dt}\left(b^{{}\dagger}_eb^{}_et\right)e^{i\left(1-\alpha\right)i\left(\nu_e-\nu_g\right) b^{{}\dagger}_eb^{}_et}d\alpha\\
	=&\, i\left(\nu_e-\nu_g\right)b^{{}\dagger}_eb^{}_ee^{i\left(\nu_e-\nu_g\right) b^{{}\dagger}_eb^{}_et}+i\left(\nu_e-\nu_g\right)t\left[-2\Gamma  b^{{}\dagger}_eb^{}_e+\sqrt{2\Gamma}\frac{e^{i(\nu_e-\nu_g)t}-1}{i(\nu_e-\nu_g)t}b^{{}\dagger}_e\mathcal{B}_\mathrm{e}^{\mathrm{in}{}}+\mathrm{h.c.}\right]\label{Eq:exp_Vib},
\end{align}
one can recast Eq.~(\ref{Eq:Polaron_E}) into
\begin{equation}
	\dot{\tilde{{\sigma}}}_e\approx -\left[i(\omega_{00}-\omega_\ell)+\gamma\right] \tilde{{\sigma}}_e-\eta_{\ell}\mathcal{S}^{{}\dagger}_e\mathcal{D}^{{}\dagger}_ee^{i\left(\nu_e-\nu_g\right) b^{{}\dagger}_eb^{}_et}\sigma^\dagger\sigma+\sqrt{2\gamma}\sigma_\mathrm{in}\mathcal{S}^{{}\dagger}_e\mathcal{D}^{{}\dagger}_ee^{i\left(\nu_e-\nu_g\right) b^{{}\dagger}_eb^{}_et},\label{Eq:EffPolaron_E}
\end{equation}
with $\tilde{{\sigma}}_e=\tilde{\tilde{\sigma}}^{}_e\exp\left[\left(i\nu_e-i\nu_g\right) b^{{}\dagger}_eb^{}_et\right]$. Here, we have dropped the second term on the right side of Eq.~(\ref{Eq:exp_Vib}) to obtain the equation above and to receive a sufficient approximation.
\label{eq:Sim_Eff_E}

Notably, Eq.~(\ref{Eq:Polaron_E})  only conclude the contribution of the vibrations projecting to the manifold of $\mathcal{P}_e$. To get the dynamics of the system in the whole Hilbert space, one need also to get the dynamics of the general polaron operator $\tilde{\tilde{\sigma}}_g=\mathcal{S}_g^\dagger\mathcal{D}_g^\dagger\sigma$ for the vibrational mode projecting to the manifold of $\mathcal{P}_g$ with $\mathcal{S}_g=\exp\left[r_s(b_g-b_g^{\dagger2})/2\right]\mathcal{P}_g$, and $\mathcal{D}_g=\left[r_d(b_g-b_g^{\dagger})\right]\mathcal{P}_g$
\begin{equation}
	\dot{\tilde{\tilde{\sigma}}}_g\approx-\left[i(\omega_{00}-\omega_\ell)+\gamma\right]\tilde{\tilde{\sigma}}_g-i(\nu_e-\nu_g)b^\dagger_gb_g\tilde{\tilde{\sigma}}_g+\eta_{\ell}\mathcal{S}_g^\dagger\mathcal{D}_g^\dagger\sigma\sigma^\dagger+\sqrt{2\gamma}\sigma_{\text{in}}\mathcal{S}_g^\dagger\mathcal{D}_g^\dagger\sigma\sigma^\dagger.
\end{equation}

Repeat the process to deriving the dynamical equation of $\tilde{{\sigma}}_e$, one can recast the above equation into
\begin{equation}
	\dot{\tilde{{\sigma}}}_g\approx  -\left[i(\omega_{00}-\omega_\ell)+\gamma\right] \tilde{{\sigma}}_g+e^{i(\nu_e-\nu_g)b^\dagger_gb_g t}\eta_{\ell}\mathcal{S}_g^\dagger\mathcal{D}_g^\dagger\sigma\sigma^\dagger+\sqrt{2\gamma}\sigma_{\text{in}}e^{i(\nu_e-\nu_g)b^\dagger_gb_g t}\mathcal{S}_g^\dagger\mathcal{D}_g^\dagger\sigma\sigma^\dagger, \label{Eq:EffPolaron_G}
\end{equation}
with $\tilde{{\sigma}}_g=\exp\left[\left(i\nu_e-i\nu_g\right) b^{\dagger}_gb_gt\right]\tilde{\tilde{\sigma}}_g$.

Meanwhile, one could also get the dynamical equation for the population of the excited state given by
\begin{equation}
	\frac{d}{dt}\sigma^{\dagger}\sigma=-2\gamma\sigma^{\dagger}\sigma+\eta_{\ell}(\sigma+\sigma^{\dagger})+\sqrt{2\gamma}(\sigma^{\dagger}\sigma_{\mathrm{in}}+\sigma\sigma_{\mathrm{in}}^{\dagger})\label{eq:Pop_E}.
\end{equation}

	\section{Vibrational Dynamics}
	\label{App:VibDy}
	As discussed in the previous section, the electronic transition is dressed by vibrations. For the further calculation of the electronic transition, we here analyze the properties of vibrations and derive the expression for the two time-correlation terms for the product of squeezing and displacement operators $\mathcal{S}_g^\dagger\mathcal{D}_g^\dagger$, $\mathcal{S}_e^\dagger\mathcal{D}_e^\dagger$.
	
	\subsection*{Nonlinear Vibrational Dynamics.}
	
	%%%%%%%%%%%%%%%%%%%%%%%%%
	%%Figure 3
	%%%%%%%%%%%%%%%%%%%%%%%%%
	\begin{figure}[thb]
		\centering
		\includegraphics[width=0.90\columnwidth]{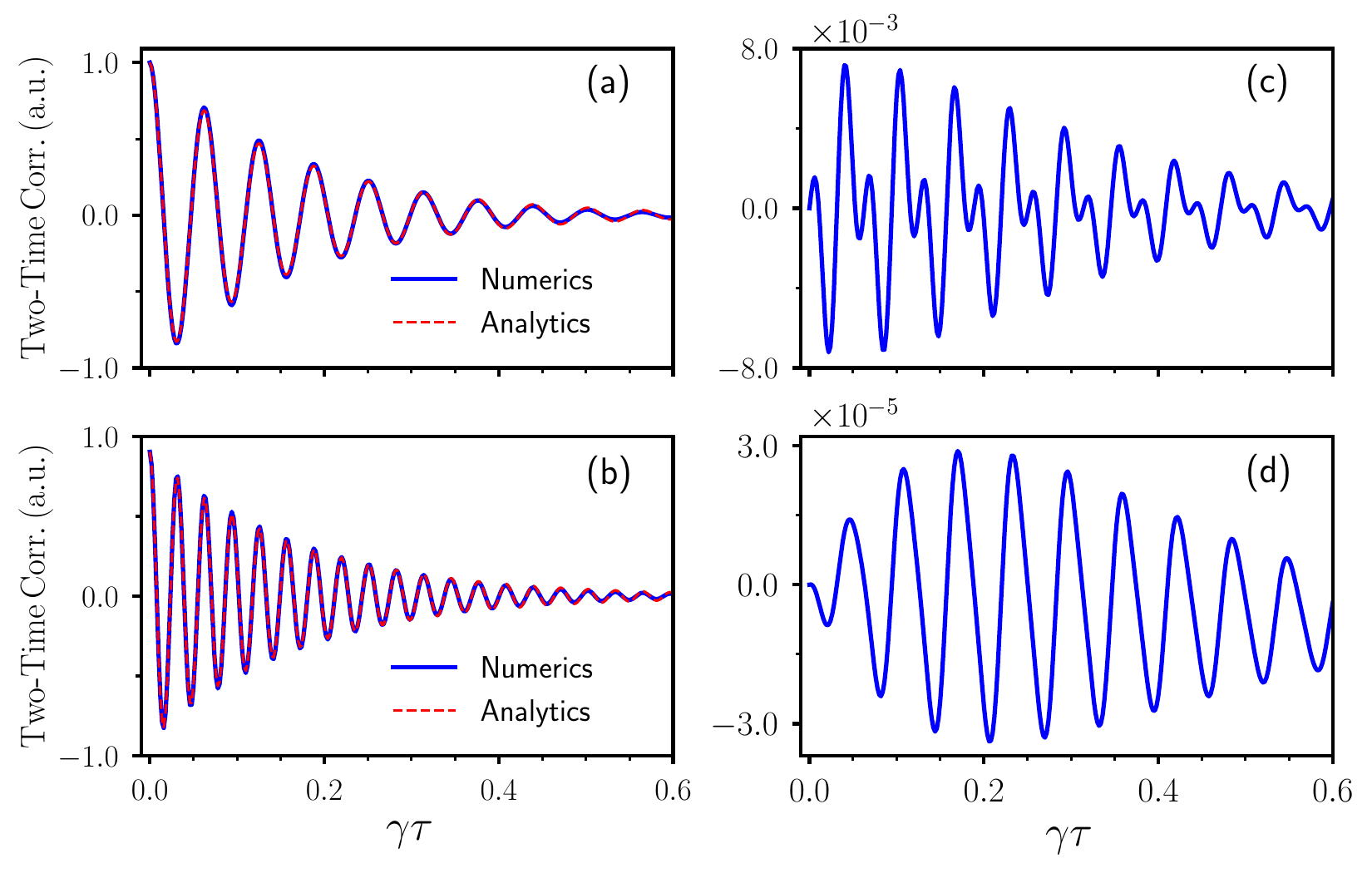}
		\caption{The two-time correlation function in the steady state for (a) $\braket{b_g(\tau)b_g^\dagger(0)}$, (b) $\braket{b_e(\tau)b_e^{\dagger}(0)}$, (c) $\braket{b_e(\tau)b_g^\dagger(0)}$,  and  $\braket{b_g(\tau)b_e^{\dagger}(0)}$ normalized in a unit factor. The blue-solid lines in (a-d) are generated with the toolbox QuTip \cite{johansson2012qutip} under the following parameters $\nu_g=1$, $\nu_e/\nu_g=2$, $\Gamma/\nu_g=0.1$, $\gamma/\nu_g=0.01$ and $\eta_{\ell}/\gamma=2$. The red-dashed lines in (a) and (b) are calculated via the first and the second term in the right side of Eq.~(\ref{eq:Corr_Vib}), respectively.}
		\label{figA}
	\end{figure}
	%%%%%%%%%%%%%%%%%%%%%%%%%
	%%%%%%%%%%%%%%%%%%%%%%%%%
	In the previous section, we have derived the effective Langevin equation~(\ref{eq:Eff_BE}) describing the dynamics of the vibrational mode projected onto the manifold of $\mathcal{P}_e$. One can easily obtain the exact solution of such an equation, which is given by
	\begin{equation}
		b^{}_e=\sqrt{2\Gamma}\int_{-\infty}^td\tau\, e^{-(i\nu_e+\Gamma)(t-\tau)} \mathcal{B}_e^\mathrm{in}(\tau)\mathcal{P}_e(\tau).
	\end{equation}
%with the vector $\mathbf{v}_{e}=\left(\begin{array}{cc}
%	b_{e}, & b_{e}^{\dagger}\end{array}\right)^{T}$ to be solved, $\mathbf{v}_{\mathrm{vib}\mbox{-}\mathrm{el}}=\left(\begin{array}{cc}
%	\Gamma r_d-i\nu,, & \Gamma r_d+i\nu\end{array}\right)^{T}$ describing the interaction between vibrations and electrons, $\mathbf{v}_{\mathrm{noise}}=\left(\begin{array}{cc}
%\mathcal{B}_{e}^{\mathrm{in}}, & \mathcal{B}_{e}^{\mathrm{in}}\end{array}\right)^{T}$ for noise input, and the drift matrix
%\begin{equation}
%\mathbf{M}=\left(\begin{array}{cc}
%		-i\left(1+2\lambda_{2}\right)\nu-\Gamma^{} & -i2\lambda_{2}\nu\\
%		i2\lambda_{2}\nu & i\left(1+2\lambda_{2}\right)\nu-\Gamma^{}
%	\end{array}\right).
%\end{equation}
%The solution is given by
%\begin{equation}
%	\mathbf{v}_e=e^{M(t-t_0)}\mathbf{v}_e(t_0)+\int_{t_0}^td\tau\,e^{M_e(t-\tau)}[\lambda_1 \mathbf{v}_{\mathrm{vib}\mbox{-}\mathrm{el}} \mathcal{P}_e\sigma^{\dagger}\sigma(\tau)\mathcal{P}_e+\sqrt{2\Gamma} \mathbf{v}_{\mathrm{noise}}(\tau)].
%\end{equation}
{Due to the correlation time for vibrations being much short than the electrons, the quantity for $\sigma^\dagger\sigma(\tau )$ varies little around $\sigma^\dagger\sigma(t)$. We can thus proceed via a Markov approximation with taking it out of the integral, which yields }
\begin{equation}
	b^{}_e=\sqrt{2\Gamma}\mathcal{P}_e(t)\int_{-\infty}^t d\tau\,e^{-(i\nu_e+\Gamma)(t-\tau)}\mathcal{B}_e^\mathrm{in}(\tau).
\end{equation}

With this, we can derive the two-time correlation for $t>\tau$ \begin{subequations}
\begin{align}
		\braket{b_e(t)b_e^{\dagger}(\tau)}_{\mathrm{vib}}=&e^{-(i\nu_e+\Gamma)(t-\tau)}\mathcal{P}_e(t),\label{eq:Cor_Vib_E}\\
		\braket{b_e(\tau)b_e^{\dagger}(t)}_{\mathrm{vib}}=&e^{-(-i\nu_e+\Gamma)(t-\tau)}\mathcal{P}_e(t),
\end{align}
where $\braket{\cdot}_{\mathrm{vib}}$ denotes taking the average over the degrees of freedom of the vibrational mode.
 \end{subequations}

Similar to the case for calculating $b_e$, we can obtaint the two-time correlation term for the vibrational operator $b_g$ projected onto the manifold of $\mathcal{P}_g$ as
\begin{subequations}
\begin{align}
\braket{b_g^{\phantom{\dagger}}(t)b_g^\dagger(\tau)}_{\mathrm{vib}}=& e^{-\left(i\nu_g+\Gamma\right)\left(t-\tau\right)}\sigma\sigma^\dagger(t),\label{eq:Cor_Vib_G}\\	
\braket{b_g^{\phantom{\dagger}}(\tau)b_g^\dagger(t)}_{\mathrm{vib}}=& e^{\left(i\nu_g-\Gamma\right)\left(t-\tau\right)}\sigma\sigma^\dagger(t).
\end{align}
 \end{subequations}
Thus, the two-time correlation function for the vibrational mode in the whole Hilbert space reads
\begin{align}\nonumber
	\braket{b(t+\tau)b^\dagger(t)}\approx&\braket{b_g^{\phantom{\dagger}}(\tau)b_g^\dagger(0)}+\braket{b_e^{\phantom{\dagger}}(\tau)b_e^{\dagger}(0)}\\
	\approx&\, e^{-\left(i\nu_g+\Gamma\right)\left(\tau\right)}\braket{\sigma\sigma^\dagger(t)}+e^{-\left(i\nu_e+\Gamma\right)\left(\tau\right)}\braket{\sigma^{\dagger}\sigma(t)}\label{eq:Corr_Vib}.
\end{align}
Here, the correlations between different manifolds have been properly dropped because of the much smaller value for such terms $\braket{b_e(\tau)b_g^\dagger(0)}$ and $\braket{b_g(\tau)b_e^{\dagger}(0)}$ comparing with that in the same manifold, as illustrate in Fig.~(\ref{figA}).

\subsection*{Two-time Correlation Function for the Displacement-Squeezing Operator.}

The squeezing operator can be written in a disentangled form
\begin{equation}
	\mathcal{S}=\exp\left[-\frac{1}{2}\tanh r_sb^{\dagger2}\right]\exp\left[-\ln\cosh r_s\left(b^{\dagger}b+\frac{1}{2}\right)\right]\exp\left[\frac{1}{2}\tanh r_sb^{2}\right],
\end{equation}
we can calculate the two-time correlation function for displacement-squeezing operator under the vacuum state. For instance, the two-time correlation for the manifold of $\mathcal{P}_e$ is given by
\begin{equation}
\begin{aligned}
			\left\langle\mathcal{S}_{e}^{{}\dagger}\left(\tau\right)\mathcal{D}_{e}^{{}\dagger}\left(\tau\right) \mathcal{D}_{e}^{}\left(t\right)\mathcal{S}_{e}^{}\left(t\right)\right\rangle_{\mathrm{vib}}
		=& \frac{e^{-r_d^2\alpha}}{\cosh(r_s)}\left\langle e^{-\beta b_{e}^{{} 2}\left(\tau\right)}e^{\alpha r_d  b_{e}^{}\left(\tau\right)}e^{\alpha r_d b_{e}^{{}\dagger}\left(t\right)}e^{-\beta b_{e}^{{}\dagger2}\left(t\right)}\right\rangle_{\mathrm{vib}},
\end{aligned}
\end{equation}
with $\alpha=\tanh r_s+1=2\nu_e/(\nu_e+\nu_g)
$ and $\beta=\left(\tanh r_s\right)/2$.

Using the generating function of the Hermite polynomials given by
\begin{equation}
	e^{2x\varphi-\varphi^{2}}=\sum_{n=0}^{\infty}H_{n}\left(x\right)\frac{\varphi^{n}}{n!},
\end{equation} we can expand the first and second exponential in the equation above in terms of Hermite polynomials with $x=\alpha r_d /(2\sqrt{\beta})$ and $\varphi=\sqrt{\beta} b_e$. Thus, we finally obtain the expression for the two-time correlation function under the Isserlis' theorem
\begin{equation}
	\begin{aligned}
		\left\langle\mathcal{S}_{e}^{{}\dagger}\left(\tau\right)\mathcal{D}_{e}^{{}\dagger}\left(\tau\right) \mathcal{D}_{e}^{}\left(t\right)\mathcal{S}_{e}^{}\left(t\right)\right\rangle_{\mathrm{vib}}
		&=\mathcal{F}_\mathrm{em}(-\nu_e,t-\tau)\mathcal{P}_e(t)
	\end{aligned}
\end{equation}
where
\begin{equation}
	\begin{aligned}
		\mathcal{F}_\mathrm{em}(\nu,t)=&\sum_{m=0}^{\infty}S_{m}^\mathrm{em}e^{-m\left(i\nu+\Gamma\right)t}, \\
		S_{m}^\mathrm{em}=&\frac{e^{-r_d^2\alpha}}{\cosh(r_s)}\left[H_{m}\left(\frac{\alpha r_d}{2\sqrt{\beta}}\right)\right]^2\frac{\beta^{m}}{m!}.
	\end{aligned}
\end{equation}
Of course, we can obtain the result for the following equation in the same way
\begin{align}
		\left\langle \mathcal{S}_{e}^{{}\dagger}\left(t\right)\mathcal{D}_{e}^{{}\dagger}\left(\tau\right)e^{i\left(\nu_e-\nu_g\right)b_e^{{}\dagger}b_e \tau}e^{-i\left(\nu_e-\nu_g\right)b_e^{{}\dagger}b_e t}\mathcal{D}_{e}^{{}}\left(t\right)\mathcal{S}_{e}^{{}}\left(t\right)\right\rangle_{\mathrm{vib}}
	=\mathcal{F}_\mathrm{em}\left(-\nu_g,t-\tau\right)\mathcal{P}_e(t).\label{eq:2t_Corr_SD_E}
\end{align}

The two-time correlation in the manifold of $\mathcal{P}_g$ reads
\begin{align}
	\braket{\mathcal{D}_g(t)\mathcal{S}_g(t)\mathcal{S}_g^\dagger(\tau))\mathcal{D}^\dagger_g(\tau)}_\mathrm{vib}
	=& \mathcal{F}_\mathrm{ab}(\nu_g,t-\tau)\mathcal{P}_g(t),
\end{align}
%where
%
%\begin{equation}
%	\begin{aligned}
%		\mathcal{F}_g(\nu,t)=&\sum_{m=0}^{\infty}S^g_{m}e^{m\left(-i\nu+\Gamma\right)t}, \\
%			S_m^g=&\frac{e^{r_d^2(\tanh r_s-1) }}{\cosh r_s}\left[\mathcal{H}_{m}\left(-\frac{i\alpha e^{r_s}}{2\sqrt{\beta}}\right)\right]^2\frac{\left(-\beta\right)^{m}}{m!},
%	\end{aligned}
%\label{eq:F_vG}
%\end{equation}
and also
\begin{equation}
	\braket{\mathcal{D}_g(t)\mathcal{S}_g(t)e^{-i\left(\nu_e-\nu_g\right)b_g^\dagger b_g t}e^{i\left(\nu_e-\nu_g\right)b_g^\dagger b_g \tau}\mathcal{S}_g^\dagger(\tau))\mathcal{D}^\dagger_g(\tau)}_\mathrm{vib}
	= \mathcal{F}_\mathrm{ab}(\nu_e,t-\tau)\mathcal{P}_g(t),\label{eq:2t_Corr_SD_G}
\end{equation}
where
\begin{subequations}
	\begin{align}
		\mathcal{F}_\mathrm{ab}(\nu,t)=&\sum_{m=0}^{\infty}S_{m}^\mathrm{ab}e^{-m\left(i\nu+\Gamma\right)t}, \\
		S_{m}^\mathrm{ab}=&\frac{e^{\alpha^{} r_d^2 \exp\left(2r_s\right)}}{\cosh(r_s)}\left[H_{m}\left(-\frac{i\alpha^{} r_d\exp\left(r_s\right)}{2\sqrt{\beta}}\right)\right]^2\frac{\left(-\beta\right)^{m}}{m!},
	\end{align}
\label{eq:Ab_Fact}
\end{subequations}
with $\alpha^{}=\tanh r_s-1=-2\nu_g/(\nu_e+\nu_g)$.

	\section{Stability Analysis}
	\label{App:SteadyPop}
	It is straightforward to obtain the expression for the Pauli operator in the whole Hilbert space by formally integrating Eqs.~(\ref{Eq:EffPolaron_E}) and (\ref{Eq:EffPolaron_G}),
	\begin{equation}
		\begin{aligned}
			\sigma(t)=&\int_{0}^td\tau\, \left(-\eta_{\ell}+\sqrt{2\Gamma}\sigma_\mathrm{in}\left(\tau\right)\right)e^{\left[-i\left(\omega_{00}-\omega_\ell\right)+\gamma\right]\left(t-\tau\right)} \mathcal{S}_{e}^{{}\dagger}\left(t\right)\mathcal{D}_{e}^{{}\dagger}\left(\tau\right)e^{i\left(\nu_e-\nu_g\right)b_e^{{}\dagger}b_e \tau}e^{-i\left(\nu_e-\nu_g\right)b_e^{{}\dagger}b_e t}\mathcal{D}_{e}^{{}}\left(t\right)\mathcal{S}_{e}^{{}}\left(t\right)\\
			&+\int_{0}^td\tau\, \left(\eta_{\ell}+\sqrt{2\Gamma}\sigma_\mathrm{in}\left(\tau\right)\right)e^{\left[-i\left(\omega_{00}-\omega_\ell\right)+\gamma\right]\left(t-\tau\right)}\mathcal{D}_g(t)\mathcal{S}_g(t)e^{-i\left(\nu_e-\nu_g\right)b_g^\dagger b_g t}e^{i\left(\nu_e-\nu_g\right)b_g^\dagger b_g \tau}\mathcal{S}_g^\dagger(\tau))\mathcal{D}^\dagger_g(\tau),
%			&+\sigma(0)e^{\left[-i\left(\omega_{00}-\omega_\ell\right)+\gamma\right]t} \mathcal{D}_{e}^{}\left(0\right)\mathcal{S}_{e}^{}\left(0\right)e^{-i\left(\nu_e-\nu_g\right)b_e^{{}\dagger}b_e t}\mathcal{S}_{e}^{\dagger{}}\left(t\right)\mathcal{D}_{e}^{\dagger{}}\left(t\right)\\
%			&+e^{\left[-i\left(\omega_{00}-\omega_\ell\right)+\gamma\right]\left(t\right)}\mathcal{D}_g(t)\mathcal{S}_g(t)e^{-i\left(\nu_e-\nu_g\right)b_g^\dagger b_g t}\mathcal{S}_g^\dagger(0))\mathcal{D}^\dagger_g(0)\sigma\left(0\right)
		\end{aligned}
	\end{equation}
with the initial value $\sigma(0)=0$.

	Under the assumption that the correlation time for the vibrations is much shorter than that for the electronic transition, we can treat the vibrations as a Markovian phonon bath.  By taking the average of the vibrational mode and substituting Eq.~(\ref{eq:2t_Corr_SD_E}) as well as Eq.~(\ref{eq:2t_Corr_SD_G}) into the equation above, we then obtain
	\begin{equation}
		\begin{aligned}
			\sigma(t)=&\int_0^\infty d\tau\, \left(-\eta_{\ell}+\sqrt{2\Gamma}\sigma_\mathrm{in}\left(\tau\right)\right)\mathcal{G}_\mathrm{em}\left(-\nu_g,t-\tau\right)\mathcal{P}_e\left(\tau\right)+\int_{0}^\infty d\tau\,\left(\eta_{\ell}+\sqrt{2\Gamma}\sigma_\mathrm{in}\left(\tau\right)\right)\mathcal{G}_\mathrm{ab}\left(\nu_e,t-\tau\right)\mathcal{P}_g\left(\tau\right),
		\end{aligned}
	\end{equation}
with $\mathcal{G}_Q(\nu,\tau)=\exp\left[\left(i\omega_\ell-i\omega_{00}+\gamma\right)\left(t-\tau\right)\right]\mathcal{F}_Q(\nu, t-\tau)\Theta(t-\tau)$ and $Q\in\left\{\mathrm{em},\mathrm{ab}\right\}$, where $\Theta(t)$ is the Heaviside step function.

Tracing over the electronic transition, we finally obtain the simplified formal solution for $\braket{\sigma}$ as
\begin{equation}
	\braket{\sigma(t)}=-\eta_{\ell}\int_0^\infty d\tau\,\mathcal{G}_\mathrm{em}(-\nu_g,t-\tau)P_e(\tau)+\eta_{\ell}\int_0^\infty d\tau\,\mathcal{G}_\mathrm{ab}(\nu_e,t-\tau)[1-P_e(\tau)]\label{eq:Average_Pauli},
\end{equation} with the population on the electronic excited state $P_e=\braket{\sigma^\dagger\sigma}$.

We proceed our calculation via the Laplace transformation (defined as $\xoverline{f}(s)=\int_0^\infty dt\, f(t)\exp(-st)$ for a time-dependent function $f(t)$ at $t\geq0$). Then Eq.~(\ref{eq:Average_Pauli}) can be written in the Laplace domain as
\begin{equation}
	\xoverline{\braket{\sigma}}=\frac{\eta_{\ell}}{s}\xoverline{\mathcal{G}}_\mathrm{ab}-\eta_{\ell}\xoverline{P}_e(\xoverline{{\mathcal{G}}}_\mathrm{ab}+\xoverline{{\mathcal{G}}}_\mathrm{em})\label{eq:Laplace_Pauli},
\end{equation}
where $\xoverline{\mathcal{G}}_\mathrm{em}$ and $\xoverline{\mathcal{G}}_\mathrm{ab}$ are the Laplace transform of $\mathcal{G}_\mathrm{ab}(\nu_e,t)$ and $\mathcal{G}_\mathrm{em}(-\nu_{g},t)$, expressed as
\begin{subequations}
	\begin{align}
		\xoverline{\mathcal{G}}_\mathrm{ab}=&\sum_{m=0}^{\infty}\frac{S_{m}^\mathrm{ab}}{s+m\Gamma+\gamma+i\left(\omega_{00}-\omega_\ell+m\nu_e\right)},\\
		\xoverline{\mathcal{G}}_\mathrm{em}=& \sum_{m=0}^{\infty}\frac{S_{m}^\mathrm{em}}{s+\gamma+m\Gamma+i\left(\omega_{00}-\omega_\ell-m\nu_g\right)}.
	\end{align}
\end{subequations}
Assuming that the molecule is prepared in the electronic ground state $\ket{g}$ followed by taking an average over the electronic transition on both sides of Eq.~(\ref{eq:Pop_E}), and applying the Laplace transformation, we finally obtain
\begin{equation}
	s\xoverline{P}_e=-2\gamma \xoverline{P_e}+\eta_{\ell}(\xoverline{\braket{\sigma}}+\xoverline{\braket{\sigma}}^*)\label{eq:Laplace_Pe}.
\end{equation}
Plugging Eq.~(\ref{eq:Laplace_Pe}) into Eq.~(\ref{eq:Average_Pauli}), we can get
\begin{subequations}
	\begin{align}
		\xoverline{\braket{\sigma}}=&\frac{i2\eta_{\ell}^3\mathcal{I}[\xoverline{\mathcal{G}}_\mathrm{ab} \xoverline{\mathcal{G}}_\mathrm{em}^*]+\eta_{\ell}(s+2\gamma)\xoverline{\mathcal{G}}_\mathrm{ab}}{s(2\gamma+s+2\eta_{\ell}^2\mathcal{R}[\xoverline{\mathcal{G}}_\mathrm{ab}+\xoverline{\mathcal{G}}_\mathrm{em}])},\\
		\xoverline{P}_e=&\frac{2\eta_{\ell}^2}{s(2\gamma+s+2\eta_{\ell}^2\mathcal{R}[\xoverline{\mathcal{G}}_\mathrm{ab}+\xoverline{\mathcal{G}}_\mathrm{em}])}\mathcal{R}[\xoverline{\mathcal{G}}_\mathrm{ab}],
	\end{align}
	\end{subequations}
	 where $\mathcal{I}[\cdot]$ and $\mathcal{R}[\cdot]$ denote taking the imaginary and real part, respectively. According to the final value theorem, we get the steady values
	\begin{subequations}
\begin{align}
	\braket{\sigma}_\mathrm{ss}=&\lim_{s\rightarrow0}s\xoverline{\braket{\sigma}}=\frac{i\eta_{\ell}^3\mathcal{I}[{\chi}_\mathrm{ab} {\chi}_\mathrm{em}^*]+\eta_{\ell}\gamma\chi_\mathrm{ab}}{\gamma+\eta_{\ell}^2\mathcal{R}[{\chi}_\mathrm{ab}+{\chi}_\mathrm{em}]},\\
	P_e^{\mathrm{ss}}=&\lim_{s\rightarrow0}s\xoverline{P}_e=\frac{\eta_{\ell}^2}{\gamma+\eta_{\ell}^2\mathcal{R}[{\chi}_\mathrm{ab}+{\chi}_\mathrm{em}]}\mathcal{R}[{\chi}_\mathrm{ab}],
\end{align}
\end{subequations}
with $\chi_Q=\lim\limits_{s\to 0}\xoverline{\mathcal{G}}_Q$. In the limit of weak driving $\eta_{\ell}\ll \gamma$, the equation above can be simplified to
\begin{subequations}
\begin{align}
			\braket{\sigma}_\mathrm{ss}\rightarrow&\eta_{\ell}\chi_\mathrm{ab}=\sum_{m=0}^{\infty}\frac{S_{m}^\mathrm{ab}}{m\Gamma+\gamma+i\left(\omega_{00}-\omega_\ell+m\nu_e\right)},\\
			P_e^{\mathrm{ss}}\rightarrow&\frac{\eta_{\ell}^2}{\gamma}\mathcal{R}[\chi_\mathrm{ab}]=\frac{\eta_{\ell}^2}{\gamma}\sum_{m=0}^{\infty}\frac{S_{m}^\mathrm{ab}\left(m\Gamma+\gamma\right)}{\left(m\Gamma+\gamma\right)^2+\left(\omega_{00}-\omega_\ell+m\nu_e\right)^2}.
\label{eq:SS}
\end{align}
\end{subequations}

%%%%%%%%%%%%%%%%%%%%%%%%%%%%%%%%%%%%%%%%%%%%%%%%%%%%%%%%%%%%%%%%%%%%%%%%%%%%%%%%%%%%%%
\section{Rate Equation}
\label{App:RateEquation}

For large vibrational relaxation rates $\Gamma\gg\eta_\ell$, the electronic transition is usually going from the lowest vibrational state in both electronic states $\ket{g}$ and $\ket{e}$. Thus, the motion of the population $p^e_m$ on the state $\ket{e;m_e}$ as well as the population $p_m^g$ on the state $\ket{g;m_g}$  are given phenomenology by
\begin{subequations}
\begin{align}
	&\begin{aligned}
		\partial_tp_m^e=&2\gamma_m^\uparrow p_0^g+\Gamma p_{m+1}^e-\Gamma p_m^e,\\
		\partial_tp_{m-1}^e=&2\gamma_{m-1}^\uparrow p_0^g+\Gamma p_{m}^e-\Gamma p_{m-1}^e,\\
		&\vdots\\
		\partial_tp_0^e=&2\gamma_0^\uparrow p_0^g+\Gamma p_{0}^e-2\gamma p_0^e-2\sum_{m=0}^\infty\gamma_m^{\downarrow}p_0^e ,
	\end{aligned}\\
	&\begin{aligned}
		\partial_tp_m^g=&2\gamma_m^\downarrow p_0^e+2\gamma_mp_0^e+\Gamma p_{m+1}^g-\Gamma p_m^g,\\
		\partial_tp_{m-1}^e=&2\gamma_{m-1}^\downarrow p_0^e+2\gamma_{m-1}p_0^e+\Gamma p_{m}^g-\Gamma p_{m-1}^g,\\
		&\vdots\\
		\partial_tp_0^g=2&\gamma_0^\downarrow p_e+2\gamma_0 p_0^e+\Gamma p_{0}^e-2\sum_{m=0}^\infty\gamma_m^{\uparrow}p_0^e,
	\end{aligned}
\end{align}
\end{subequations}
where $\gamma_m$ describe the incoherent spontaneous emission progress $\ket{e;0}\rightarrow\ket{g;m}$ satisfying $\sum\limits_{m=0}^{\infty}\gamma_m=\gamma$. It is obvious that the total population on the state $\ket{e}$ is the sum of all the occupations of its sublevels as $p_e=\sum\limits_{m=0}^{\infty}p_m^e$. Thus, we have
\begin{equation}
	\dot{p}_e=2\sum\limits_{m=0}^{\infty} \gamma_m^\uparrow p_0^g-2\gamma p_0^e-2\sum_{m=0}^\infty\gamma_m^{\downarrow}p_0^e.\label{eq:Pop_Rate}
\end{equation}
Considering less population on states $\ket{e;m_e>0}$ and $\ket{g;m_g>0}$ due to the large vibrational relaxation $\Gamma$, we can simplify Eq.~(\ref{eq:Pop_Rate}) into
\begin{equation}
	\dot{p}_e=2\sum\limits_{m=0}^{\infty} \gamma_m^\uparrow (1-p_e)-2\left(\gamma +\sum_{m=0}^\infty\gamma_m^{\downarrow}\right)p_e
\end{equation}
by assuming $p_e\approx p_0^e$ and  $p_g=1-p_e\approx p_0^g$.

%%%%%%%%%%%%%%%%%%%%%%%%%%%%%%%%%%%%%%%%%%%%%%%%%%%%%%%%%%%%%%%%%%%%%%%%%%%%%%%%%%%%%%
\section{Absorption and Emission Spectra}
\label{App:Ab&Em}

\subsection*{Effective Quantum Langevin Equation for the Commutator.}
In this subsection, we will describe the process to find the absorption and emission spectroscopic signal. Before that, let us introduce the following correlators ``dressed''  by a vibrational mode
\begin{equation}
	\tilde{C}_\mathrm{em}^e(t+\tau)=\sigma^\dagger(t)\sigma(t+\tau)\mathcal{S}^{{}\dagger}_e(t+\tau)\mathcal{D}^{{}\dagger}_e(t+\tau)\quad\mathrm{and} \quad
	\tilde{C}_\mathrm{em}^g(t+\tau)=\mathcal{S}_g^\dagger(t+\tau)\mathcal{D}_g^\dagger(t+\tau)\sigma^\dagger(t)\sigma(t+\tau)
\end{equation}
for emission and
\begin{equation}
		\tilde{C}_\mathrm{ab}^e(t+\tau)=\sigma(t+\tau)\sigma^\dagger(t)\mathcal{S}^{{}\dagger}_e(t+\tau)\mathcal{D}^{{}\dagger}_e(t+\tau)\quad\mathrm{and}\quad
	\tilde{C}_\mathrm{ab}^g(t+\tau)=\mathcal{S}_g^\dagger(t+\tau)\mathcal{D}_g^\dagger(t+\tau)\sigma(t+\tau)\sigma^\dagger(t)
\end{equation}
for absorption.
According to the discussion  in the Subsec.~(\ref{sub:ET-EQLE}), one will obtain
\begin{subequations}
	\begin{align}
		&\begin{aligned}
			\frac{d{\tilde{C}}_\mathrm{em}^e}{d\tau}\approx&-\left(i\omega_{00}+\gamma\right)\tilde{C}_\mathrm{em}^e-i(\nu_e-\nu_g)\tilde{C}_\mathrm{em}^e  b^{{}\dagger}_eb_e(t+\tau)-\eta_{\ell} e^{-i\omega  t}\sigma^\dagger(t)\mathcal{S}^{{}\dagger}_e(t+\tau)\mathcal{D}^{{}\dagger}_e(t+\tau)\mathcal{P}_e(t+\tau)\\&+\sqrt{2\gamma}\sigma^\dagger(t)\sigma_\mathrm{in}(t+\tau)\mathcal{S}^{{}\dagger}_e(t+\tau)\mathcal{D}^{{}\dagger}_e(t+\tau)\mathcal{P}_e(t+\tau),
		\end{aligned}\\
	&\begin{aligned}
		\frac{d{\tilde{C}}_\mathrm{em}^g}{d\tau}\approx&-\left(i\omega_{00}+\gamma\right)\tilde{C}_\mathrm{em}^g-i(\nu_e-\nu_g)b^\dagger_gb_g(t+\tau)\tilde{C}_\mathrm{em}^e
		+\eta_{\ell} e^{-i\omega  t}\mathcal{S}_g^\dagger(t+\tau)\mathcal{D}_g^\dagger(t+\tau)\mathcal{P}_g(t+\tau)\sigma^\dagger(t)\\
		&+\sqrt{2\gamma}\mathcal{S}_g^\dagger(t+\tau)\mathcal{D}_g^\dagger(t+\tau)\mathcal{P}_e(t+\tau)\sigma^\dagger(t)\sigma_\mathrm{in}(t+\tau),
	\end{aligned}\\
		&\begin{aligned}
		\frac{d{\tilde{C}}_\mathrm{ab}^e}{d\tau}\approx&-\left(i\omega_{00}+\gamma\right)\tilde{C}_\mathrm{ab}^e-i(\nu_e-\nu_g)\tilde{C}_\mathrm{ab}^e  b^{{}\dagger}_eb_e(t+\tau)
		-\eta_{\ell} e^{-i\omega  t}\sigma^\dagger(t)\mathcal{S}^{{}\dagger}_e(t+\tau)\mathcal{D}^{{}\dagger}_e(t+\tau)\mathcal{P}_e(t+\tau)\\&+\sqrt{2\gamma}\sigma_\mathrm{in}(t+\tau)\sigma^\dagger(t)\mathcal{S}^{{}\dagger}_e(t+\tau)\mathcal{D}^{{}\dagger}_e(t+\tau)\mathcal{P}_e(t+\tau),
	\end{aligned}\\
	&\begin{aligned}
	\frac{d{\tilde{C}}_\mathrm{ab}^g}{d\tau}\approx&-\left(i\omega_{00}+\gamma\right)\tilde{C}_\mathrm{ab}^g-i(\nu_e-\nu_g)b^\dagger_gb_g(t+\tau)\tilde{C}_\mathrm{ab}^e
	+\eta_{\ell} e^{-i\omega  t}\mathcal{S}_g^\dagger(t+\tau)\mathcal{D}_g^\dagger(t+\tau)\mathcal{P}_g(t+\tau)\sigma^\dagger(t)\\&+\sqrt{2\gamma}\mathcal{S}_g^\dagger(t+\tau)\mathcal{D}_g^\dagger(t+\tau)\mathcal{P}_e(t+\tau)\sigma_\mathrm{in}(t+\tau)\sigma^\dagger(t).
\end{aligned}
	\end{align}
\label{eq:Commutator}
\end{subequations}

\subsection*{Emission Spectra in the Transient Regime.}
We now assume the molecule is initially prepared in the excited state $\ket{e;\tilde{0}}$ to compute the spectrum of emission in the transient regime. By setting  $\eta_{\ell}=0$, one could obtain the expression of the two-time correlation function $\braket{\sigma^\dagger(0)\sigma(\tau)}$ through Eq.~(\ref{eq:Commutator})
\begin{equation}
	\braket{\sigma^\dagger(0)\sigma(t)}=\mathcal{F}_\mathrm{em}(\nu_g,\tau)e^{-\left(i\omega_{00}+\gamma\right) t}.
\end{equation}
Taking the Fourier transformation gives the expression of the emission spectrum
\begin{equation}
	S_\mathrm{Em}(\omega)=2\mathcal{R}\int_{0}^\infty d\tau\,\mathcal{F}_\mathrm{em}(\nu_g,\tau)e^{\left[i\omega-\left(i\omega_{00}+\gamma\right)\right] \tau}=\sum_{m=0}^{\infty}\frac{S_{m}^\mathrm{em}\left(\gamma+m\Gamma\right)}{\left(\gamma+m\Gamma\right)^2+\left(\omega_{00}-\omega-m\nu_g\right)^2}.
\end{equation}

\subsection*{Absorption Spectra in the Stability Regime.}

The formal solution of the two-time correlation function $\braket{\sigma(t+\tau)\sigma^\dagger(t)}$ is given
\begin{equation}
	\begin{aligned}
		\braket{\sigma(t+\tau)\sigma^\dagger(t)}=&-\eta_{\ell}\int_{t}^\infty d\tau^{} \mathcal{G}_\mathrm{em}(-\nu_g,t+\tau-\tau^{})\braket{\mathcal{P}_e\left(\tau^{}\right)\sigma^\dagger(t)}
		+\eta_{\ell}\int_t^\infty d\tau \mathcal{G}_\mathrm{ab}(\nu_e,t+\tau-\tau^{})\braket{\mathcal{P}_g\left(\tau^{}\right)\sigma^\dagger(t)}\\&+\mathcal{P}_g(t)\mathcal{F}_\mathrm{ab}(\nu_e,\tau)e^{-\left(i\omega_{00}+\gamma\right) t}.
	\end{aligned}
\end{equation}

For simplicity, let us assume the amplitude for the driving field is very weak, so that only few molecules are occupy their excited state, i.e., $\mathcal{P}_e(t)\ll1$ and $\mathcal{P}_g(t)\approx 1$. The solution will become
\begin{equation}
	\begin{aligned}
		\braket{\sigma(t+\tau)\sigma^\dagger(t)}\approx&\,\eta_{\ell}\int_t^\infty d\tau^{}\, \mathcal{G}_\mathrm{ab}(\nu_e,t+\tau-\tau^{})\braket{\sigma^\dagger(t)}+\mathcal{F}_\mathrm{ab}(\nu_e,\tau)e^{-\left(i\omega_{00}+\gamma\right) \tau}\\
		=&\,\eta_{\ell} \int_t^\infty d\tau^{}\,e^{-(i\omega_{00}-i\omega_\ell+\gamma)(t+\tau-\tau^{})}\mathcal{F}_\mathrm{ab}(\nu_e,t+\tau-\tau^{})\Theta(t+\tau-\tau^{})\braket{\sigma^\dagger(t)}+\mathcal{F}_\mathrm{ab}(\nu_e,\tau)e^{-\left(i\omega_{00}+\gamma\right) \tau}.
	\end{aligned}
\label{eq:Comm_Ab}
\end{equation}

Let us pay attention to the steady state regime by setting $t\rightarrow\infty$.  According to Eq.~(\ref{eq:SS}), we can get the expectation value of the transition  dipole moment beyond the rotating frame
\begin{equation}
	\lim\limits_{t\to \infty}\braket{\sigma(t)}=\chi_\alpha e^{-i\omega_\ell t}.
\end{equation}
%now write the Pauli operator $\sigma(t)$ as the sum of its expect value $\braket{\sigma}(t)$ and the fluctuation part $\delta \sigma(t)$ as $\sigma=\braket{\sigma}+\delta\sigma$.   Thus, the auto two-time correlation for the fluctuation operator $\delta\sigma(t)$ is then expressed as
%\begin{equation}
%	\begin{aligned}
%		\braket{\delta\sigma(t+\tau)\delta\sigma^\dagger(t)}=\braket{\delta\sigma(t+\tau)\delta\sigma^\dagger(t)}=\braket{\sigma(t+\tau)\sigma^\dagger(t)}-\braket{\sigma(t+\tau)}\braket{\sigma(t)}.
%	\end{aligned}\label{eq:Comm_Flu}
%\end{equation}
Inserting the expression of the function $\mathcal{F}_\mathrm{ab}(\nu,t)$ given by Eq.~(\ref{eq:Ab_Fact}) into  Eq.~(\ref{eq:Comm_Ab}), we obtain
\begin{subequations}
	\begin{align}
		\lim\limits_{t\to \infty}\braket{\sigma(t+\tau)\sigma^\dagger(t)}\approx
			& \sum_{m, n=0}^{\infty} \frac{\eta_{\ell}^2 S_m^\mathrm{ab} S_n^\mathrm{ab}\left[e^{-i \omega_\ell \tau}-e^{-\left(i \omega_{00}+i m \nu_e+\gamma+m \Gamma\right) \tau}\right]}{\left[\gamma+m \Gamma+i\left(\omega_{00}+m \nu-\omega_\ell\right)\right]\left[\gamma+n \Gamma-i\left(\omega_{00}+n \nu_e-\omega_\ell\right)\right]}\label{eq:<Sigma>^2}\\
			&+ \sum_{m=0}^{\infty} S_m^\mathrm{ab} e^{-\left(i \omega_{0 0}+i m \nu_e+\gamma+m \Gamma\right) \tau}.
	\end{align}
\end{subequations}
  As $\eta_{\ell}\ll \gamma$, the contribution of the term in Eq.~(\ref{eq:<Sigma>^2}) can be ignored. Finally, we can simplify the expression of  $\braket{\sigma(t+\tau)\sigma^\dagger(t)}$  into
  \begin{equation}
  	\lim\limits_{t\to \infty}\braket{\sigma(t+\tau)\sigma^\dagger(t)} =\sum_{m=0}^{\infty} S_m^\mathrm{ab} e^{-\left(i \omega_{0 0}-i m \nu_e+\gamma+m \Gamma\right) \tau}.
  \end{equation}
  Performing the Fourier transform, the absorption spectrum is obtained as
  \begin{equation}
  	S_{\mathrm{Ab}}(\omega)=2\mathcal{R}\int_{0}^\infty d\tau\,	\lim\limits_{t\to \infty}\braket{\sigma(t+\tau)\sigma^\dagger(t)} e^{i\omega \tau}=\sum_{m=0}^{\infty}\frac{S_{m}^\mathrm{ab}\left(\gamma+m\Gamma\right)}{\left(\gamma+m\Gamma\right)^2+\left(\omega+m\nu_e-\omega_{00}\right)^2}.
  \end{equation}

\end{document}